**Ejecta fate and deposition in the Didymos–Dimorphos system after the DART impact**

Kosuke Kurosawa[1,2,3] and Satoshi Takada[4]

**Abstract**

Kinetic impact is a promising technique for asteroid deflection in planetary defense. The NASA Double Asteroid Redirection Test (DART) spacecraft collided with Dimorphos, the satellite of the binary asteroid (65803) Didymos, at 6.1449 km $s^{-1}$ in 2022. Observations revealed changes in both the orbit of Dimorphos around Didymos and the heliocentric orbit of the Didymos–Dimorphos system. The ESA Hera spacecraft will arrive at the system at the end of 2026 to investigate the post-impact surfaces. Here, we present an analytical model, based on Maxwell's $Z$ model, to predict ejecta behavior following the DART impact. The model provides the relationships among ejection velocity, ejection angle, and ejecta mass, allowing the momentum enhancement factor $\beta$ to be estimated for both Dimorphos and the Didymos–Dimorphos system. We show that the observed $\beta$ can be reproduced when 0.3–1% of the impact kinetic energy is converted into ejecta kinetic energy and the effective strength of the target is $<10^2$ Pa. The escaping mass is estimated to be 0.3–3.0% of the mass of Dimorphos. The globally averaged and minimum ejecta thicknesses on Dimorphos are estimated to be ~2.0 m and ~0.5 m, respectively, indicating that the entire surface was resurfaced by ejecta deposition, although boulders larger than ~1 m were only partially buried. The model also predicts the momentum partition parameter $\kappa_p$ to be as low as ~0.6, suggesting that the mass and the bulk density of Dimorphos may be revised upward. These predictions provide quantitative constraints for interpreting forthcoming Hera observations. (250/250 words)

[1] Corresponding author kosuke.kurosawa@people.kobe-u.ac.jp

[2] Department of Human Environmental Science, Graduate School of Human Development and Environment, Kobe University, 3–11, Tsurukabuto, Nada-ku, Kobe, Hyogo 657-8501, Japan

[3] Planetary Exploration Research Center, Chiba Institute of Technology, 2-17-1, Tsudanuma, Narashino, Chiba 275-0016, Japan

[4] Institute of Engineering, Tokyo University of Agriculture and Technology, 2-24-16, Nakacho, Koganei, Tokyo 184-8588, Japan

## 1. INTRODUCTION

A bolide over Chelyabinsk Oblast in Russia in 2013 caused the largest airburst on Earth since the well-known Tunguska event in 1908 (Popova et al., 2013). The meteoroid is estimated to have been ~20 m in diameter prior to atmospheric entry. The frequency of airburst events caused by bolides of a similar size is approximately once every 30–100 yr (e.g., Bland and Artemieva, 2003), which is the same order of magnitude as a typical human lifetime. This event demonstrated that bolide impacts should be recognized as natural hazards and highlighted the need for planetary defense research.

Ahrens and Harris (1992) proposed that kinetic impactors can deflect objects of ~100 m in diameter from Earth-crossing trajectories. The Asteroid Impact & Deflection Assessment (AIDA) collaboration is ongoing. AIDA involves a kinetic impactor and a follow-up asteroid rendezvous spacecraft. The NASA Double Asteroid Redirection Test (DART) spacecraft successfully collided with the asteroid Dimorphos, which is a satellite of the binary asteroid (65803) Didymos, at a speed of 6.1449 km $s^{-1}$ on 26 September 2022 (Daly et al., 2023). A large amount of ejecta was observed by LICIACube, which separated from the DART spacecraft before impact, as well as by the James Webb Space Telescope (JWST), the Hubble Space Telescope (HST), and ground-based telescopes (Graykowski et al., 2023; Jewitt et al., 2023; Li et al., 2023; Dotto et al., 2024; Hirabayashi et al., 2025; Farnham et al., 2025). The continuous ejecta exhibited two distinct velocity components (“fast” and “slow”) and discrete ejected boulders were observed (Graykowski et al., 2023; Jewitt et al., 2023; Farnham et al., 2025). The impact shortened the orbital period of Dimorphos by approximately 33 minutes (Thomas et al., 2023; Naidu et al., 2024). The ESA Hera spacecraft was launched in 2024 as a follow-up mission and is scheduled to begin its detailed investigation of the Didymos–Dimorphos system in late 2026 (e.g., Michel et al., 2022).

The degree of orbital deflection due to hypervelocity impacts has been characterized by the momentum enhancement factor $\beta$, which is defined as the ratio of the total momentum transferred to the target body to the incident momentum of the spacecraft. Given that impact ejecta carry away both mass and momentum from the target body, the recoil of the ejecta also contributes to the momentum transfer. The simplest definition of $\beta$ is as follows (e.g., Cheng et al., 2016):

$$\beta = 1 + \frac{p_{\mathrm{ej},\parallel}}{M_{\mathrm{proj}} v_{\mathrm{imp}}}, \quad (1)$$

where $p_{\mathrm{ej},\parallel}$ is the momentum of the ejecta parallel to the impactor trajectory, $M_{\mathrm{proj}}$ is the mass of the kinetic impactor, and $v_{\mathrm{imp}}$ is the impact velocity. A more general definition of $\beta$ that accounts for three-dimensional impact geometries can be found in previous studies (e.g., Rivkin et al., 2021; Makadia et al., 2026). The $\beta$ value for the DART impact has been estimated to be >2, based on the change in the orbital period of Dimorphos after the impact (Cheng et al., 2023). A value of $\beta > 2$ indicates that the ejecta recoil contributed more momentum than the impactor itself. To maximize the scientific return from the DART and Hera missions, it is essential to understand the consequences of the DART impact and to predict the post-impact surface evolution of both asteroids to support the interpretation of Hera observations.

An analytical expression for $\beta$ has been proposed (Holsapple and Housen, 2012; Cheng et al., 2016) based on point-source theory (Holsapple and Schmidt, 1982). The high $\beta$ value estimated for the DART impact may reflect the physical properties of the target asteroid, because impact ejection processes depend on gravity and mechanical properties. However, the analytical expression cannot be applied directly to the DART impact because it assumes a flat target surface, an assumption that becomes particularly problematic when impact ejection is dominated by gravity rather than material strength. For example, assuming a flat target and using the saturated-soil parameters of Schmidt and Housen (1987), which are representative of a cohesionless target, the transient crater radius is estimated to be 190 m under gravity-dominated conditions. In contrast, the radius of Dimorphos $R_{\mathrm{Di}}$ is only 75 m (Daly et al., 2023).

Shock physics modeling has been a powerful tool for predicting the outcomes of the DART impact (e.g., Raducan et al., 2019, 2024; Raducan and Jutzi, 2022; Stickle et al., 2022; Jiao et al., 2024; Senel et al. 2025) because it can account for the effects of target curvature on impact processes. However, this approach, particularly in three dimensions, is computationally expensive for the following reasons. According to Pierazzo et al. (2008), at least 20 computational cells per projectile radius are needed to accurately resolve the peak shock-pressure distribution using grid-based hydrocodes. Given that the size ratio of Dimorphos to the DART spacecraft is ~$10^2$, approximately $(2 \times 20 \times 10^2)^3 \sim 10^{10}$–$10^{11}$ computational grid cells would be required in three dimensions to resolve the shock-pressure distribution throughout the target at the recommended spatial resolution. Even in two dimensions, ~$10^7$ grid cells would be required. The computational challenge becomes more severe in the gravity-dominated regime, even in two-dimensional simulations. The characteristic time for crater formation $t_{\mathrm{cr}}$ can be roughly estimated as $t_{\mathrm{cr}} \sim (2R_{\mathrm{Di}}/g)^{1/2}$ (e.g., Melosh, 1989, P. 77) $\sim 2 \times 10^3$ s, where $g$ = 5 μG is the gravitational acceleration on the surface of Dimorphos. At the recommended spatial resolution of 20 computational cells per projectile radius, the Courant–Friedrichs–Lewy (CFL) condition gives a characteristic time step of $\Delta t \sim R_{\mathrm{DART}}/(20C_{\mathrm{s}}) \sim 1 \times 10^{-5}$ s, where $R_{\mathrm{DART}} \sim 0.5$ m is the effective radius of the DART spacecraft and $C_{\mathrm{s}} \sim 3 \times 10^3$ m s$^{-1}$ is the assumed bulk sound speed of the target material. At the escape speed of Dimorphos (~9 cm s$^{-1}$), ejecta would travel only ~1 μm during each time step. Consequently, maintaining this time step throughout crater formation would require approximately $10^8$ integration steps. This illustrates the computational difficulty of exploring a broad parameter space using conventional shock physics simulations. It should be mentioned here that Jutzi et al. (2022) developed a fast time-integration method to reduce this computational cost. Their approach switches from an equation of state (EOS) representing the initial target material to a modified EOS with a reduced effective sound speed after shock passage, thereby allowing larger time steps under the CFL condition. Nevertheless, compressional waves in granular materials can still propagate much faster than the escape speed on Dimorphos (e.g., Yasui et al., 2015). Thus, resolving the crater excavation and ejection of very slow material remains computationally demanding even with accelerated time-integration schemes.

The heliocentric deflection of the Didymos–Dimorphos system has been directly detected through detailed observations of its orbit following the DART impact (Makadia et al., 2026). By combining the measured heliocentric deflection with the observed change in Dimorphos's orbit around Didymos, Makadia et al. (2026) estimated the mass of Dimorphos using a formulation involving the momentum partition parameter $\kappa_p$, defined as the ratio of the momentum carried by ejecta escaping the binary system to that carried by ejecta escaping Dimorphos, projected along their respective orbital along-track directions. The estimated mass was $(2.7 \pm 0.4) \times 10^9$ kg, corresponding to a bulk density of 1540 $\pm$ 220 kg m$^{-3}$ when combined with the volume derived from the shape model of Dimorphos (Daly et al., 2024). Both values are substantially lower than the previous estimates of $4.3 \times 10^9$ kg and 2400 kg m$^{-3}$ (Daly et al., 2023).

Although Makadia et al. (2026) assumed that $\kappa_p \sim$ U[0.9, 1.0], where U[a, b] denotes a uniform distribution between a and b, they also examined the sensitivity of the estimated Dimorphos mass to $\kappa_p$ by varying the lower bound of this distribution from 0 to 0.9. They further noted that $\kappa_p$ should be constrained by Hera observations. The uncertainty in $\kappa_p$ reflects, in part, our limited understanding of how ejecta momentum is partitioned between material escaping the Didymos–Dimorphos system and material remaining gravitationally bound to it. Because this partitioning is strongly influenced by ejecta with velocities near the system escape velocity, accurately characterizing the low-speed ejecta population is essential for constraining $\kappa_p$.

In this study, we propose a simple model to predict the consequences of the DART impact, based on a modified version of Maxwell's *Z* model (Maxwell, 1977; Kurosawa and Takada, 2019). This model predicts the relationships among ejection position, velocity, angle, and mass of the ejecta without being constrained by the CFL condition. It enables the investigation of ejecta with velocities much lower than the escape velocity of Dimorphos and is therefore particularly suitable for determining the ultimate fate of impact ejecta on and around Dimorphos. This capability is essential for interpreting the post-impact state that will be observed by Hera and for constraining the momentum partition parameter $\kappa_p$.

## 2. MODEL DESCRIPTION

We extend the modified Maxwell's *Z* model of Kurosawa and Takada (2019) by incorporating the effects of target curvature into the excavation flow. The *Z* model provides streamlines describing the excavation flow, whose geometry allows us to approximately evaluate the effects of target curvature.

We define a streamtube as the volume enclosed by two adjacent streamlines. We briefly describe the basic concepts, key equations, and limitations of the original version of the *Z* model in section 2.1. We describe the procedure for introducing target curvature into the *Z* model in section 2.2. We derive the volume elements along the streamlines and calculate the mass and the work done against gravity and material strength within each streamtube. In section 2.3, we briefly describe the residual velocity concept proposed by Melosh (1985). Kurosawa and Takada (2019) combined the *Z* model and the residual velocity concept to overcome the limitations of the original *Z* model. This approach allows us to calculate the total kinetic energy of the material within a given streamtube. We describe the calculation procedure used to determine the relationships among launch position, velocity, angle, and mass in section 2.4. Finally, we summarize the calculation conditions in section 2.5.

### 2.1. Original version of Maxwell's *Z* model

Maxwell (1977) constructed an analytical model to determine the geometry of a cratering flow field in two-dimensional polar coordinates ($r$, $\theta$), where $r$ is the radial distance from the impact point and $\theta$ is the polar angle measured from the vertically downward direction. The direction $\theta = 0$ is vertically downward. The radial component of the particle velocity $u_{p,r}$ was assumed as follows:

$$u_{p,r} = \alpha(t)r^{-Z}, \quad (2)$$

where $\alpha(t)$ is the time-dependent amplitude of the flow and $Z$ is a decay exponent. Given that the excavation flow is described as an incompressible flow (i.e., $\nabla\cdot\boldsymbol{u} = 0$), the geometry of a streamline at a given horizontal distance $R$ is expressed, regardless of the details of $\alpha(t)$, as follows:

$$r = R(1 - \cos\theta)^{\frac{1}{Z-2}}. \quad (3)$$

Because α(t) is unspecified, the original *Z* model predicts the streamline geometry but cannot uniquely determine the excavation velocity or the termination of crater growth. The inherent limitations of Maxwell's *Z* model are briefly discussed in section 4.1. Nevertheless, we adopted this model as the basis for developing a computationally inexpensive approach to estimating ejecta behavior on small asteroids.

### 2.2. Incorporating curvature effects into the *Z* model

In this section, we describe the procedure for introducing target curvature into the *Z* model. We define the angle $\theta_c$, which is as the polar angle at which a streamline intersects the surface of Dimorphos, as a function of horizontal distance from the impact point $R$. The angle $\theta_c$ satisfies the following equation:

$$r = R(1 - \cos\theta_c)^{\frac{1}{Z-2}} = 2R_{Di}\cos\theta_c, \quad (4)$$

where $R_{Di}$ = 75 m is the radius of Dimorphos. Equation (4) is solved numerically for $\theta_c$ at each value of $R$. The volume element of an axisymmetric excavation streamtube is expressed in terms of the streamline coordinates ($R$, $\theta$) as follows:

$$\mathrm{d}V = 2\pi R^2(1 - \cos\theta)^{\frac{3}{Z-2}}\sin\theta \mathrm{d}R\mathrm{d}\theta. \quad (5)$$

The derivation of Eq. (5) is given in Appendix A. The effects of target curvature can be incorporated by setting the upper limit of the volume integral to $\theta = \theta_c$. The mass $m_{exc}$, gravitational energy $E_{grav}$, and strength energy $E_{str}$ pertaining to each streamtube at $R$ can be calculated as follows:

$$m_{exc} = \int_0^{\theta_c} \rho \mathrm{d}V, \quad (6)$$

$$E_{grav} = \int_0^{\theta_c} \left(1 - \frac{H}{2R_{Di}}\right) g_{surf} H \rho \mathrm{d}V, \quad (7)$$

and

$$E_{str} = \int_0^{\theta_c} \varepsilon Y \mathrm{d}V, \quad (8)$$

where $\rho$ is the density of the material composing Dimorphos, $g_{surf}$ = 5 μG is the gravitational acceleration on the surface of Dimorphos, $H$ is the depth measured from the surface of Dimorphos at the zenith angle to $r$ measured from the center of Dimorphos, $\varepsilon$ is the volumetric strain, and $Y$ is a constant that describes the strength of the material on Dimorphos. The depth $H$ along the streamline can be calculated from a simple geometric consideration (Appendix B). If (6)–(8) are integrated from $\theta = 0$ to $\pi/2$, we can apply the model to flat targets. We assumed that $\rho$ is a constant and that $\varepsilon$ is unity in this study. The gravitational energy required for excavation $E_{grav}$ is evaluated using the gravitational potential of a uniform-density sphere, accounting for the decrease in gravitational acceleration with depth. The energy loss associated with material strength is approximated as $E_{str} = \varepsilon Y V_{exc}$, where $V_{exc}$ is the total excavated volume. Because these parameters appear only through their product, the model is sensitive to the effective strength $\varepsilon Y$, rather than to $\varepsilon$ and $Y$ independently. We adopted $\varepsilon = 1$ as an order-of-magnitude approximation, and variations in $\varepsilon$ can be equivalently represented by rescaling $Y$. Thus, the strength values inferred from our model should be interpreted in terms of the assumed effective deformation parameter. The main assumptions of our model are that the streamline geometry is unaffected by target curvature and that material within the streamtubes is ballistically ejected from the curved surface along the streamlines. Figure 1 shows three examples of streamlines, which intersect the same launch position. The ejection behavior is illustrated by three arrows. The ejection angles depend on the $Z$ value used in our model. Figure 2 graphically shows the definition of $\theta_c$ and two ejection angles $\theta_{ej}$ and $\psi_{ej}$, which are measured from horizontal and tangential lines to the launch position on the surface of Dimorphos, respectively. We calculate the ejecta momentum component in Equation (1) using $\theta_{ej}$ to calculate $\beta$. The local ejection angle $\psi_{ej}$ was used to roughly estimate the ejecta depositional processes by assuming ballistic flight. We used the same model as in Kurosawa et al. (2019). An analytical formula for the ballistic range on curved surfaces (e.g., Melosh, 1989, p. 87) was implemented in the model. This formula accounts for the variation of gravitational acceleration in a spherically symmetric gravitational field.

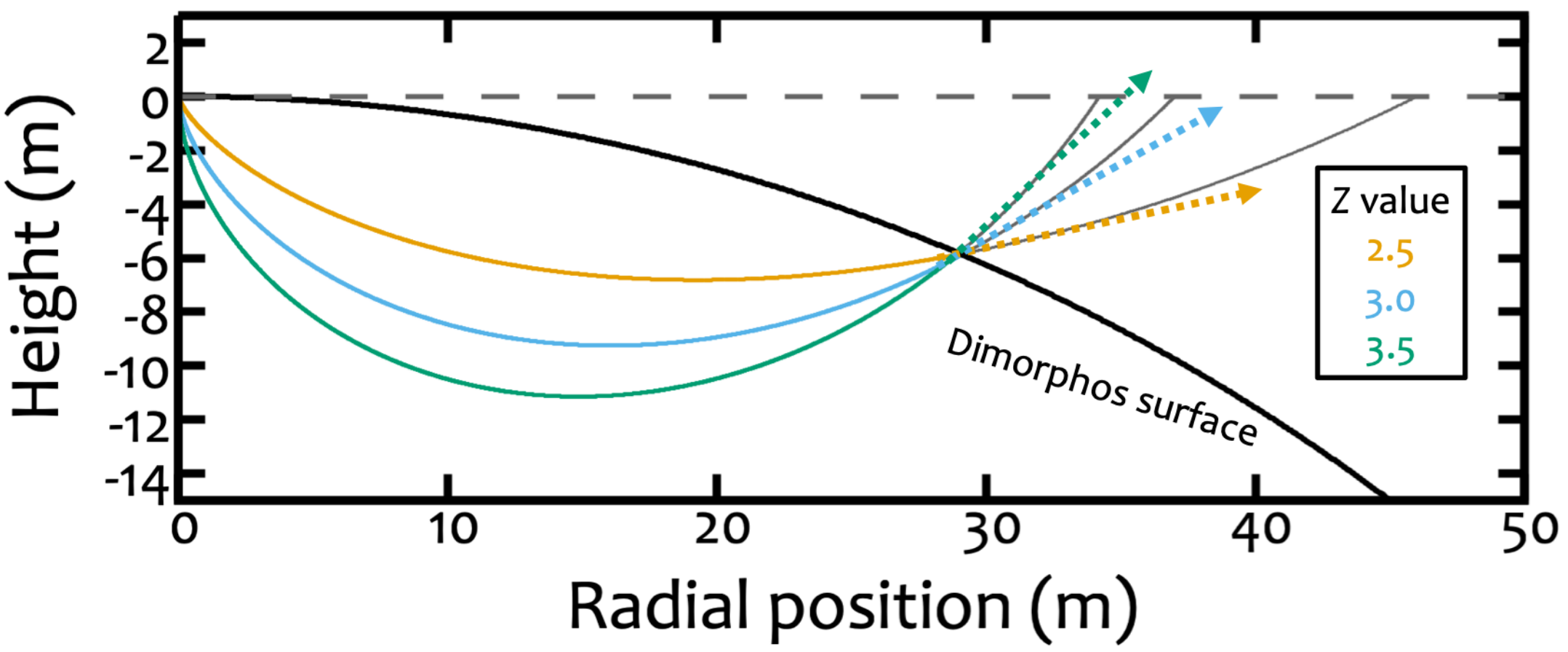

**Figure 1.** Examples of streamlines calculated with the $Z$ model. We used $Z = 2.5$, 3.0, and 3.5 as shown in the legend. We assumed that the material within the streamtube is ballistically ejected along the streamline at the surface of Dimorphos as shown by the three arrows. The surface of Dimorphos and a horizontal line (dashed gray line) are also shown.

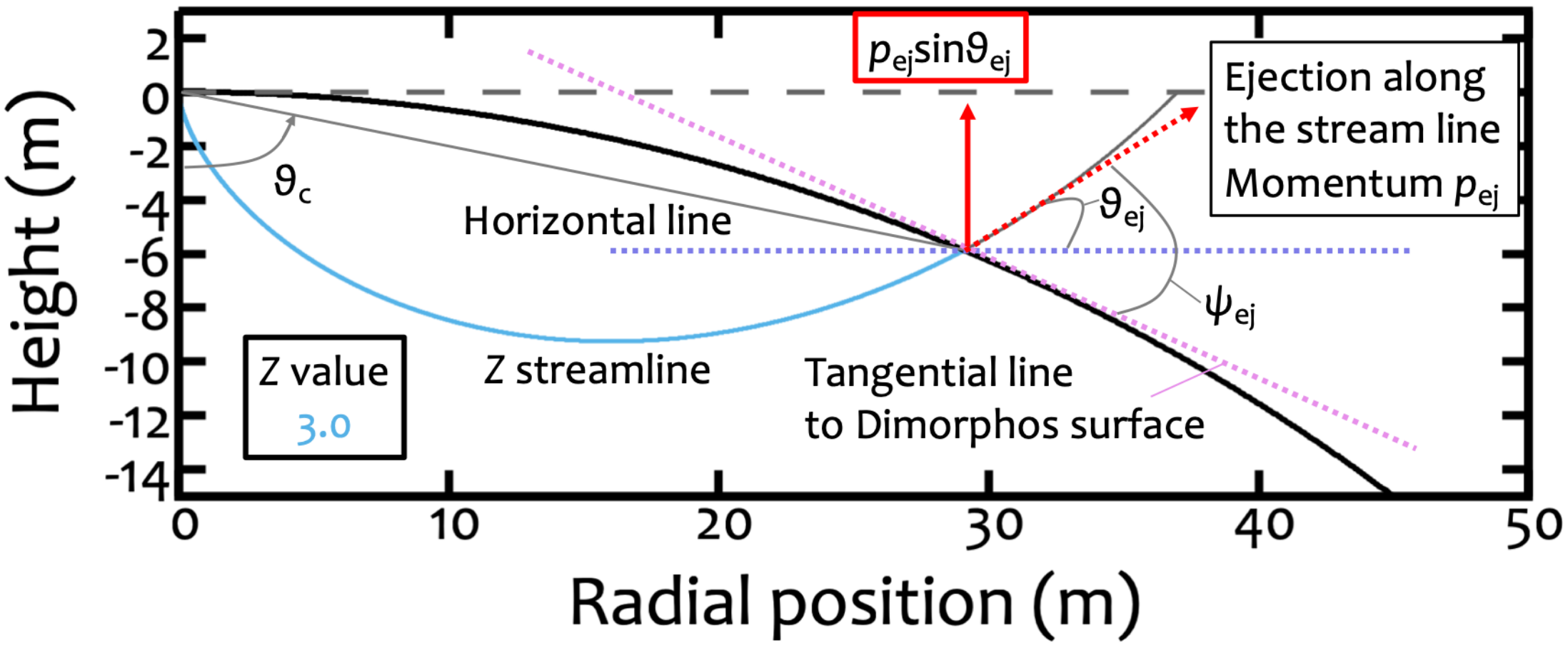

**Figure 2.** As for Fig. 1, except that the definitions of the angles required for calculating the $\beta$ values and ejecta thickness distribution are shown. We used $Z = 3.0$ in this figure, The vertical component of the momentum of the ejecta $p_{ej}\sin\theta_{ej}$ corresponds to $p_{ej,||}$ used in (1) (see the main text).

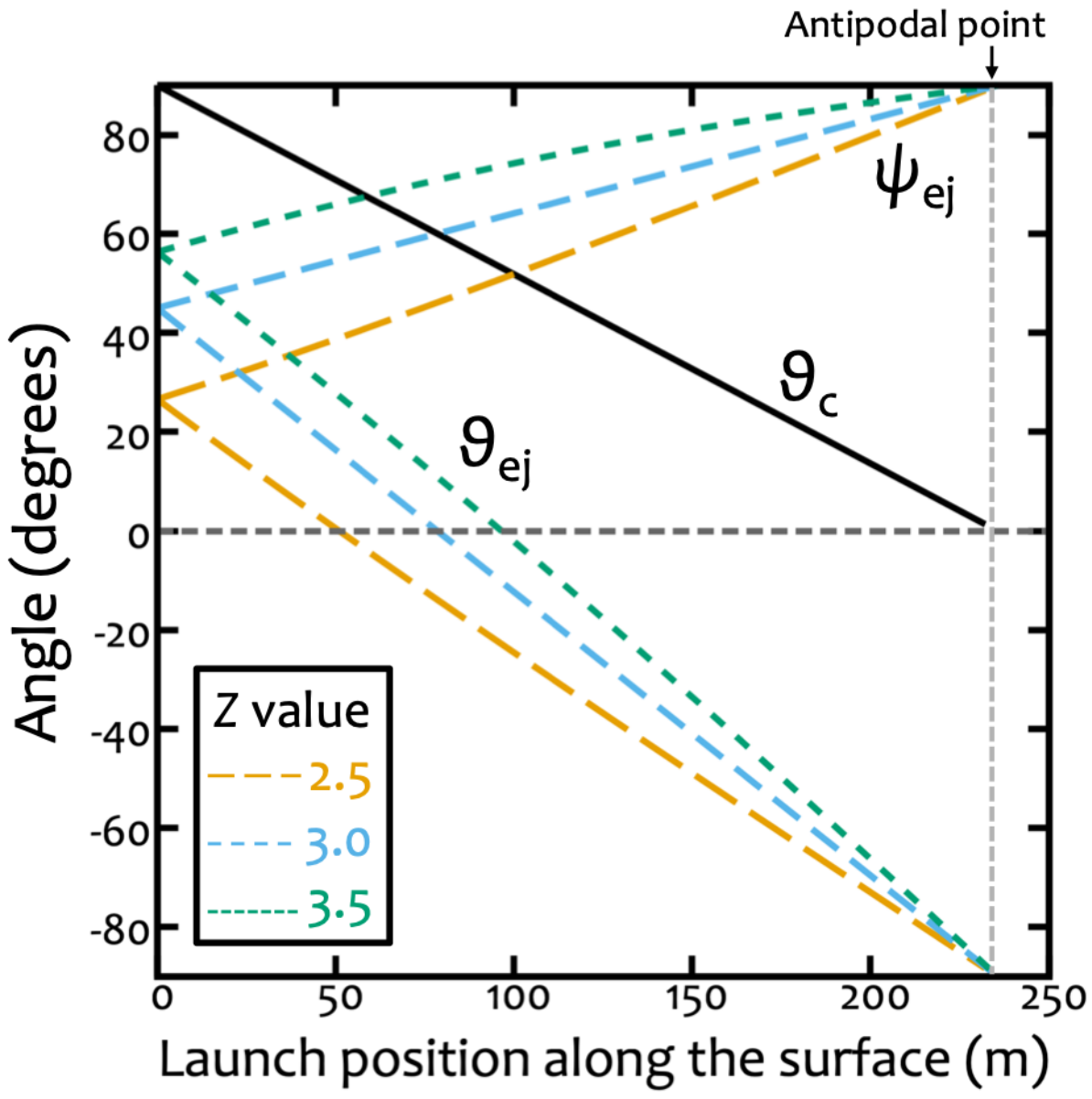

**Figure 3.** The three angles used in our model as a function of distance from the launch position along the surface of Dimorphos. We used $Z = 2.5$, 3.0, and 3.5. Note that the meaning of the zero values (the horizontal gray dashed line) differs among the three angles because the reference directions of $\theta_c$, $\theta_{ej}$, and $\psi_{ej}$ are vertically downward, horizontal (parallel to the tangent plane at the impact point), and tangent to the local surface of Dimorphos at the launch position, respectively, as illustrated in Fig. 2.

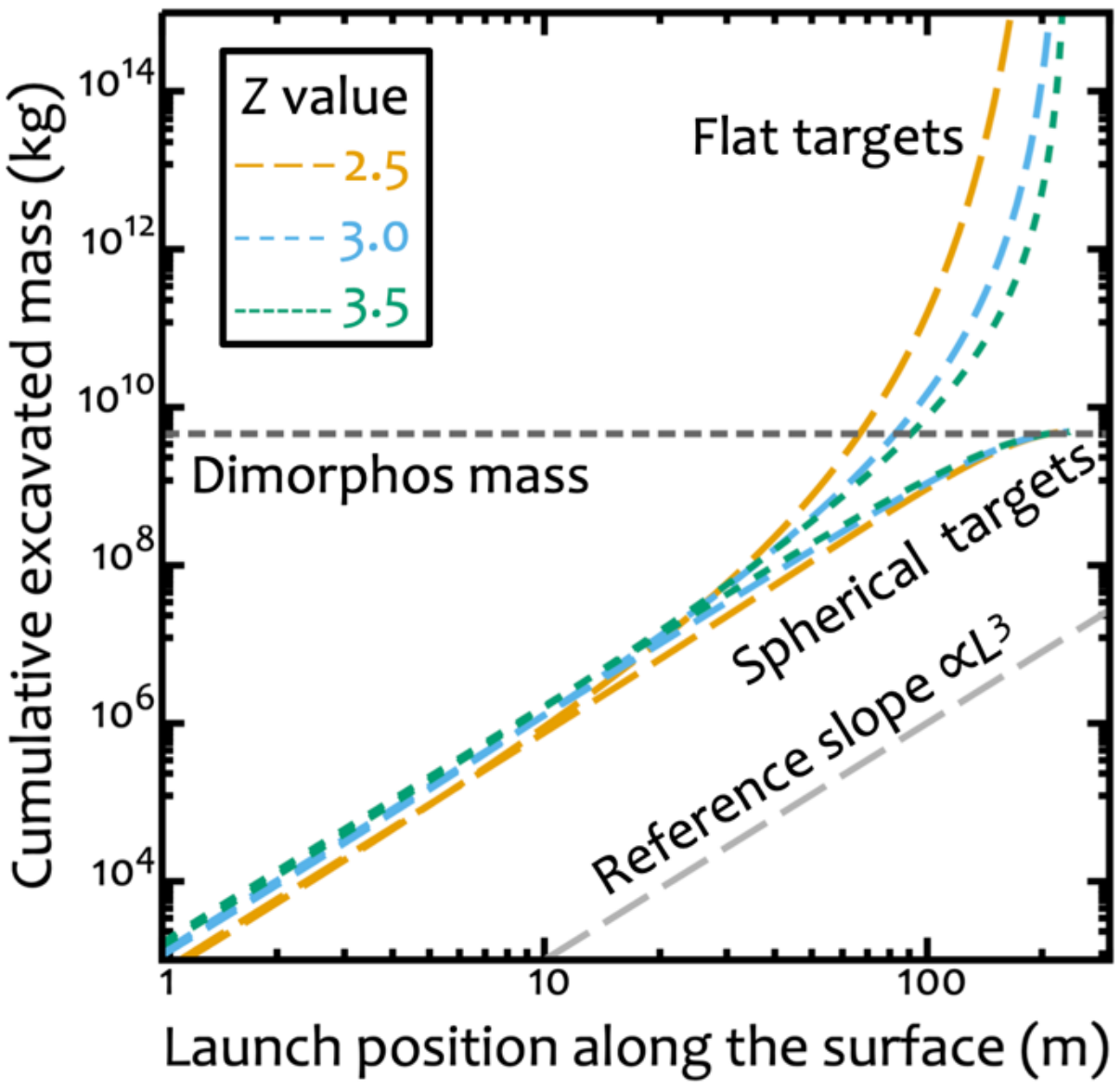

**Figure 4.** Cumulative excavated mass $M_{\mathrm{exc}}$ integrated from the impact point to a given launch position as a function of the distance of the launch position from the impact point along the surface of Dimorphos $L$. We used $Z = 2.5$, 3.0, and 3.5 as in Figs 1 and 3. The reference line for $M_{\mathrm{exc}} \propto L^3$ is also shown (gray dashed line).

Figure 3 shows $\theta_c$, $\theta_{\mathrm{ej}}$, and $\psi_{\mathrm{ej}}$ as functions of the distance along the surface from the impact point to the launch position $L_{\mathrm{ej}}$. A distance of ~240 m corresponds to the antipodal point of the impact position. $\theta_c$ and $\theta_{\mathrm{ej}}$ monotonically decrease with increasing distance, and $\psi_{\mathrm{ej}}$ exhibits the opposite trend. Figure 4 shows the cumulative excavated mass $M_{\mathrm{exc}}$ as a function of $L_{\mathrm{ej}}$. We calculated $M_{\mathrm{exc}}$ by summing the masses of the streamtubes obtained from (6). For reference, $M_{\mathrm{exc}}$ in the case of a flat target is also plotted. The $M_{\mathrm{exc}}$ values for spherical targets gradually deviate from those for flat targets at $L_{\mathrm{ej}} \sim 10$ m and asymptotically approach the mass of the body.

### 2.3. Residual velocity concept and kinetic energy carried by the impact ejecta

Melosh (1985) proposed that an excavation flow is caused by an entropy gain due to irreversible heating. The residual velocity $u_{\mathrm{p,res}}$ after a shock-release cycle can be calculated by integrating the Riemann invariant along an isentrope from the shocked state. Kurosawa and Takada (2019) systematically evaluated the relationship between the particle velocity immediately after the arrival of shock waves $u_{\mathrm{p,max}}$ and the residual velocity with the Tillotson equation of state (Tillotson, 1962), as follows:

$$u_{\mathrm{p,res}} = C u_{\mathrm{p,max}}^{m}, \quad (9)$$

where $C$ is a constant and $m$ is an exponent. The values of $C$ and $m$ are listed in table 1 of Kurosawa and Takada (2019). The distribution of $u_{\mathrm{p,max}}$ is expressed (e.g., Croft, 1982; Pierazzo et al., 1997) as follows:

$$u_{\mathrm{p,max}} = \begin{cases} u_{\mathrm{p0}} & (r \le R_{\mathrm{ic}}) \\ u_{\mathrm{p0}} \left(\frac{r}{R_{\mathrm{ic}}}\right)^{-n} & (r > R_{\mathrm{ic}}) \end{cases}, \quad (10)$$

where $u_{\mathrm{p0}}$ is the $u_{\mathrm{p,max}}$ within the isobaric core, $R_{\mathrm{ic}}$ is the radius of the isobaric core, and $n$ is a decay exponent. Legacy experiments with nuclear explosions show that the exponent $n$ is 1.87 for both igneous and sedimentary rocks (Perret and Bass, 1975). By combining (9)–(10), $u_{\mathrm{p,res}}$ is given as follows:

$$u_{\mathrm{p,res}} = \begin{cases} C u_{\mathrm{p0}}^{m} & (r \le R_{\mathrm{ic}}) \\ C u_{\mathrm{p0}}^{m} \left(\frac{r}{R_{\mathrm{ic}}}\right)^{-mn} & (R_{\mathrm{ic}} < r < R_{\mathrm{th}}) \end{cases}, \quad (11)$$

where $R_{\mathrm{th}}$ is the threshold radius, which corresponds to the transition position from the shockwave to an elasto-plastic wave (e.g., Melosh, 1989) where $u_{\mathrm{p,max}}$ ($R_{\mathrm{th}}$) = 0.8 km s$^{-1}$ in the case of granite (Kurosawa & Takada, 2019). Note that $u_{\mathrm{p,res}} = 0$ is assumed at $r > R_{\mathrm{th}}$ because the entropy gain is rather small at a location far from the impact point. By assuming a self-similar and hemispherical shock propagation, the total energy partitioned into the excavation flow $E_{\mathrm{exc}}$ is given as follows:

$$E_{\mathrm{exc}} = 2\pi\rho \int_{0}^{R_{\mathrm{th}}} \frac{u_{\mathrm{p,res}}^{2}}{2} r^{2} \mathrm{d}r. \quad (12)$$

Since $R_{\mathrm{th}}$ is at most a few times the projectile radius $R_{\mathrm{proj}}$ at relatively low impact velocities, we can neglect the effects of the target curvature in the calculation of $E_{\mathrm{exc}}$. The total available energy for impact ejection $E_{\mathrm{ej0}}$ can be estimated based on a geometrical consideration as follows:

$$E_{\mathrm{ej0}} = \left(\frac{Z-2}{Z+1}\right) E_{\mathrm{exc}}. \quad (13)$$

The coefficient $\left(\frac{Z-2}{Z+1}\right)$ corresponds to the ratio of the excavated volume to the entire volume of the crater cavity (e.g., Croft, 1980). We assume that the same ratio applies to the partitioning of excavation-flow energy. Thus, only the fraction of $E_{\mathrm{exc}}$ is assigned to the ejected material in our model.

We assumed that $E_{ej0}$ is initially deposited at the impact point and $E_{ej0}$ is expended gradually with increasing $R$. The available energy for impact ejection at a given $R$ is defined as $E_{ej}(R)$. We defined an "energy density" associated with the impact ejection $e_{ej}$ as a function of $R$ as follows. Figure 5 shows a schematic diagram of the region where the kinetic energy pertaining to the impact ejection has spread with time. The region is defined as the red shaded area in Fig. 5, which is the area bounded by the surface of Dimorphos and a hemisphere centered at the impact point and passing through the intersection of the streamline with the surface at a given $R$. Given that the mass within the region is defined as $M_{hem}$, $e_{ej}$ is estimated as follows:

$$e_{ej}(R) = \frac{E_{ej}(R)}{M_{hem}(R)}. \quad (14)$$

The derivation of $M_{hem}(R)$ can be found in Appendix C. We can calculate $M_{hem}$ at each integration in the $\theta$ direction. The kinetic energy $E_{kin}$ associated with the streamtube at a given $R$ is given by

$$E_{kin}(R) = e_{ej}(R) m_{exc}(R), \quad (15)$$

where $m_{exc}(R)$ is the mass of the material contained in the excavation streamtube between $R$ and $(R+\mathrm{d}R)$, where $\mathrm{d}R$ is the radial width of the streamtube. $E_{ej}(R)$ is expressed as follows:

$$E_{ej}(R) = E_{ej0} - \int_0^R E_{kin}(R')\mathrm{d}R'. \quad (16)$$

Consequently, we can calculate the balance between the kinetic energy $E_{kin}$ and the required work against gravity and strength ($E_{grav} + E_{str}$) at a given $R$ with (7)–(8) and (16).

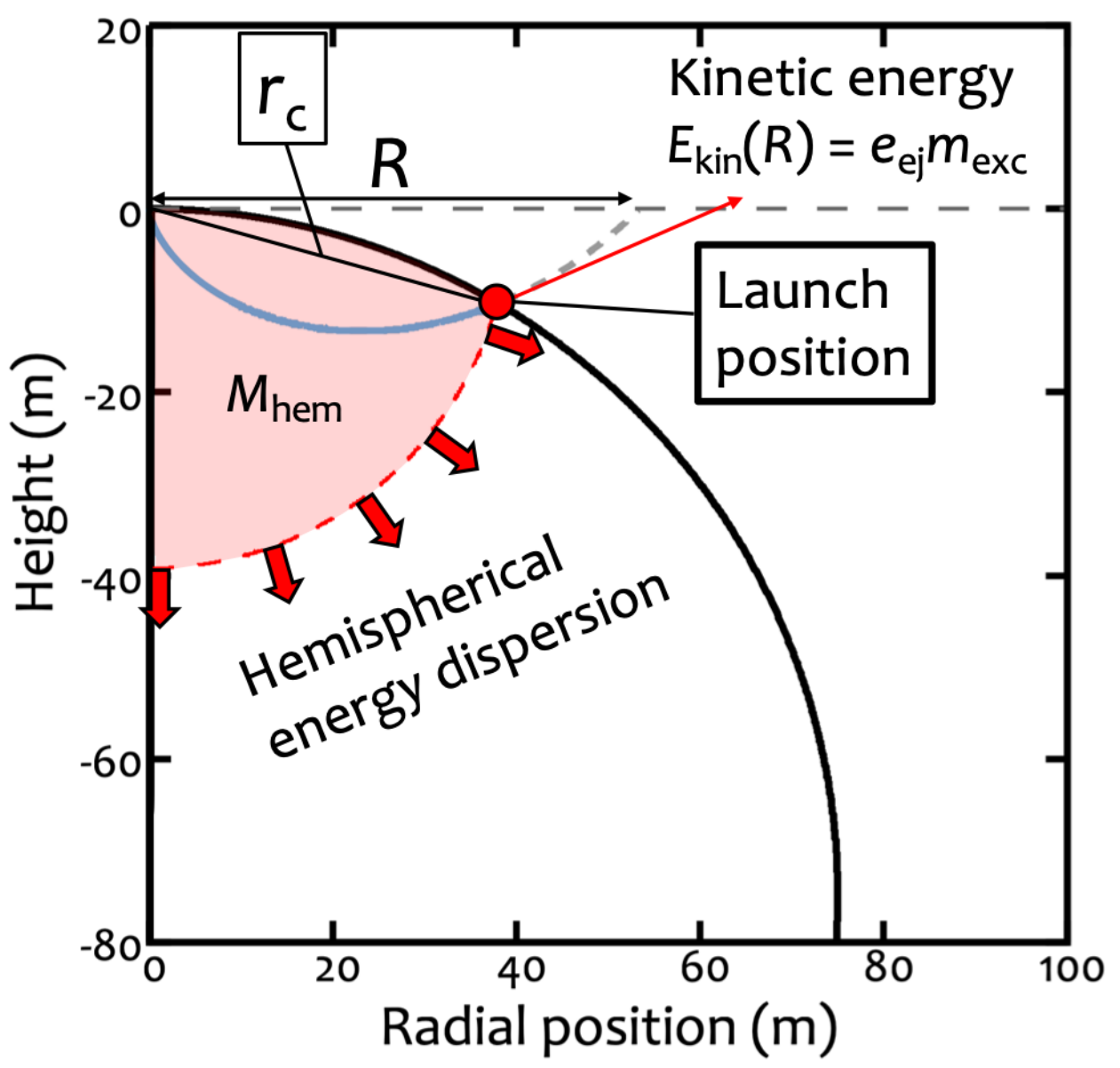

**Figure 5.** Schematic diagram of the procedure for calculating the kinetic energy pertaining to each streamtube. A hemispherical energy dispersion in a spherical body is shown as the red shaded region and red arrows.

### 2.4. Calculation of the momentum enhancement factor $\beta$ and the ejecta momentum ratio $\kappa_p$

If $E_{kin} > (E_{grav} + E_{str})$, the material in the streamtube is ejected. The ejection velocity $v_{ej}(R)$ is calculated from the energy conservation as follows:

$$v_{ej}(R) = \sqrt{\frac{2(E_{kin} - E_{grav} - E_{str})}{m_{exc}}}. \quad (17)$$

Here, $v_{ej}(R)$ represents the magnitude of the ejection velocity, whereas the ejection direction is determined separately from the geometry of the excavation streamlines. The gravitational energy $E_{grav}$ is calculated from the depth of the excavated material below the spherical surface of Dimorphos. Therefore, Eq. (17) does not require the ejecta to be launched normal to the target surface.

Crater growth ceases when $E_{kin} = (E_{grav} + E_{str})$. As such, we can estimate the relationship between the launch position, ejection velocity, ejection angle, and excavated mass associated with each streamtube. We integrated the component of the ejecta momentum opposite to the impact direction with respect to $R$ until crater growth ceased. Finally, we can calculate $\beta(v_{ej} > v)$ using (1) as

$$\beta(v_{ej} > v) = 1 + \frac{\int_{\{R:\, v_{ej} > v\}} v_{ej}(R) \sin\theta_{ej}(R) \frac{dM_{exc}(R)}{dR} dR}{M_{proj} v_{imp}}, \quad (18)$$

where $M_{exc}(R)$ is the cumulative excavated mass from the impact point to the radial distance $R$. Note that $v_{ej}$ decreases monotonically with increasing $R$. The resulting $\beta(v_{ej} > v_{esc})$ can be compared with the observed $\beta$ value (Cheng et al., 2023), where $v_{esc}$ = 8.7 cm s$^{-1}$ is the escape velocity from Dimorphos. We calculated the ejecta momentum ratio $\kappa_p$ introduced by Makadia et al. (2026) as

$$\kappa_p = \frac{\beta(v_{ej} > v_{esc,DD}) - 1}{\beta(v_{ej} > v_{esc}) - 1}, \quad (19)$$

where $v_{esc,DD}$ = 24.5 cm s$^{-1}$ (Raducan and Jutzi, 2022) is the escape velocity from the Didymos–Dimorphos system. In our model, $\kappa_p$ is the momentum carried by ejecta escaping the Didymos–Dimorphos system relative to that carried by ejecta escaping Dimorphos. It should be mentioned here that the $\kappa_p$ value calculated by our model represents an estimate based on the ejecta velocities immediately after impact and does not account for the subsequent orbital evolution of the ejecta under the gravitational influence of Didymos and Dimorphos and/or solar radiation pressure.

Although our model also requires numerical integrations in two dimensions (i.e., $R$ and $\theta$), the integrations are not limited by the CFL condition. Consequently, our model can be used as a quick-look tool to predict the ejecta behavior following the DART impact. Based on the resolution convergence test presented in Appendix D, we adopted integration step sizes of $0.3R_{proj}$ and $9 \times 10^{-3}$ degrees in the radial and angular directions, respectively.

## 2.5. Calculation conditions

The calculation settings are summarized in Table 1. The impact velocity $v_{imp}$ and gravitational acceleration $g_{surf}$ were fixed at the same values as the DART impact. We considered a solid aluminum projectile with a radius of 37 cm. The projectile mass and kinetic energy were the same as for the DART spacecraft. We used the parameters pertaining to granite to calculate $E_{ej0}$. The granite parameters (Allen, 1967) were chosen because a set of Tillotson equation-of-state parameters is available for this material, allowing us to calculate the shock response and residual particle velocity required by our model. The adopted density (2630 kg m$^{-3}$) is within the uncertainty range of the density of the Didymos–Dimorphos system reported by Daly et al. (2023), although it is higher than the bulk density of Dimorphos estimated by Makadia et al. (2026). Thus, the use of granite should not be interpreted as an assumption that Dimorphos has the composition or bulk density of granite. Note that the current β value reported by Cheng et al. (2023) is expressed as a range due to uncertainties in the density of Dimorphos and the propagation of observational uncertainties. Since ($\beta$ – 1) depends linearly on the assumed density for a fixed ejecta velocity distribution, a value calculated for a given density can readily be rescaled to another assumed density.

Although Dimorphos is an oblate ellipsoid (e.g., Daly et al., 2024), we approximated it as a sphere with a radius of 75 m, corresponding to its volume-equivalent radius (Daly et al., 2023). Because the DART spacecraft impacted near the equator, the local surface curvature is greater in the north–south direction and smaller in the east–west direction than that of a perfect sphere. Thus, the results obtained with the spherical approximation should be regarded as first-order estimates. Hirabayashi et al. (2025) also investigated the effects of surface curvature on the ejecta curtain using Maxwell's $Z$ model. While their study focused on the geometry of the ejecta curtain, our study incorporates target curvature into the estimation of excavated mass, energy, and ejecta velocities. The two studies therefore address complementary aspects of curvature effects on impact ejecta.

The Maxwell $Z$ values were varied from 2.1 to 4.0, because the range covers reasonable values to describe a cratering flow field (e.g., Croft, 1980). The constant strength was varied over eight orders of magnitude to constrain the strength of the surface material on Dimorphos. Since the radius of the isobaric core $R_{ic}$ is one of the parameters used to calculate $E_{ej0}$ in (11)–(13), we introduced a free parameter $\gamma$ with values ranging from 1 to 2, and $R_{ic}$ is expressed as $R_{ic} = \gamma R_{proj}$. Note that the upper limit of $\gamma$ is $3^{1/3} \approx 1.44$ (Senshu et al., 2002) in a strict sense. Although $\gamma$ originates from the isobaric-core formulation, it is treated here as a free parameter controlling the excavation-flow energy. Accordingly, values up to $\gamma = 2$ are explored as a phenomenological extension rather than as a strict geometric estimate of the isobaric-core radius. The maximum particle velocity at the impact point $u_{p0}$ used in (10)–(11) is calculated to be 3.3 km $s^{-1}$ for the impact conditions adopted in this study. Throughout our calculations, we assume a collision between nonporous, homogeneous bodies: an aluminum projectile and a granite target. In reality, the DART spacecraft had a complex, nonspherical structure (e.g., Stickle et al., 2022), and Dimorphos is likely heterogeneous (e.g., Raducan et al., 2024; Senel et al., 2025; Farnham et al., 2025). The grain density of Dimorphos also remains uncertain. We therefore treat $\gamma$ as a free parameter to explore uncertainties in the initial excavation-flow energy $E_{ej0}$. Variations in $\gamma$ can also represent, to some extent, uncertainties in the energy partitioning from the impact kinetic energy to the excavation flow, including those associated with the values of $u_{p0}$ and $R_{th}$ in (11).

Figure 6 shows $E_{ej0}$ as a function of $\gamma$. Results for different $Z$ values are also shown. To explore a wide parameter space, the total available energy for impact ejection was assumed to be 0.01%–1% of the kinetic energy of the DART spacecraft. As such, we only considered the ejecta that were launched at a relatively low velocity.

**Table 1.** Calculation settings.

| | |
|---|---|
| Impact velocity $v_{imp}$ (km s$^{-1}$) | 6.1449 (same as the DART spacecraft) |
| Gravitational acceleration $g$ (G)[1] | $5 \times 10^{-6}$ |
| Projectile material | aluminum (density = 2700 kg m$^{-3}$) |
| Projectile radius $R_{proj}$ (m) | 0.37 |
| Target material | granite (density = 2630 kg m$^{-3}$) |
| Target radius $R_t$ (m)[1] | 75 (Dimorphos), $7.5 \times 10^7$ (flat target) |
| Maxwell $Z$ value | 2.1–4.0 |
| Constant strength $Y$ (Pa) | $10^{-4}$ to $10^4$ |
| Radius of the isobaric core γ (projectile radius) | 1.0–2.0 |

[1] Note that we set the target radius to $7.5 \times 10^7$ m to approximate a flat target.

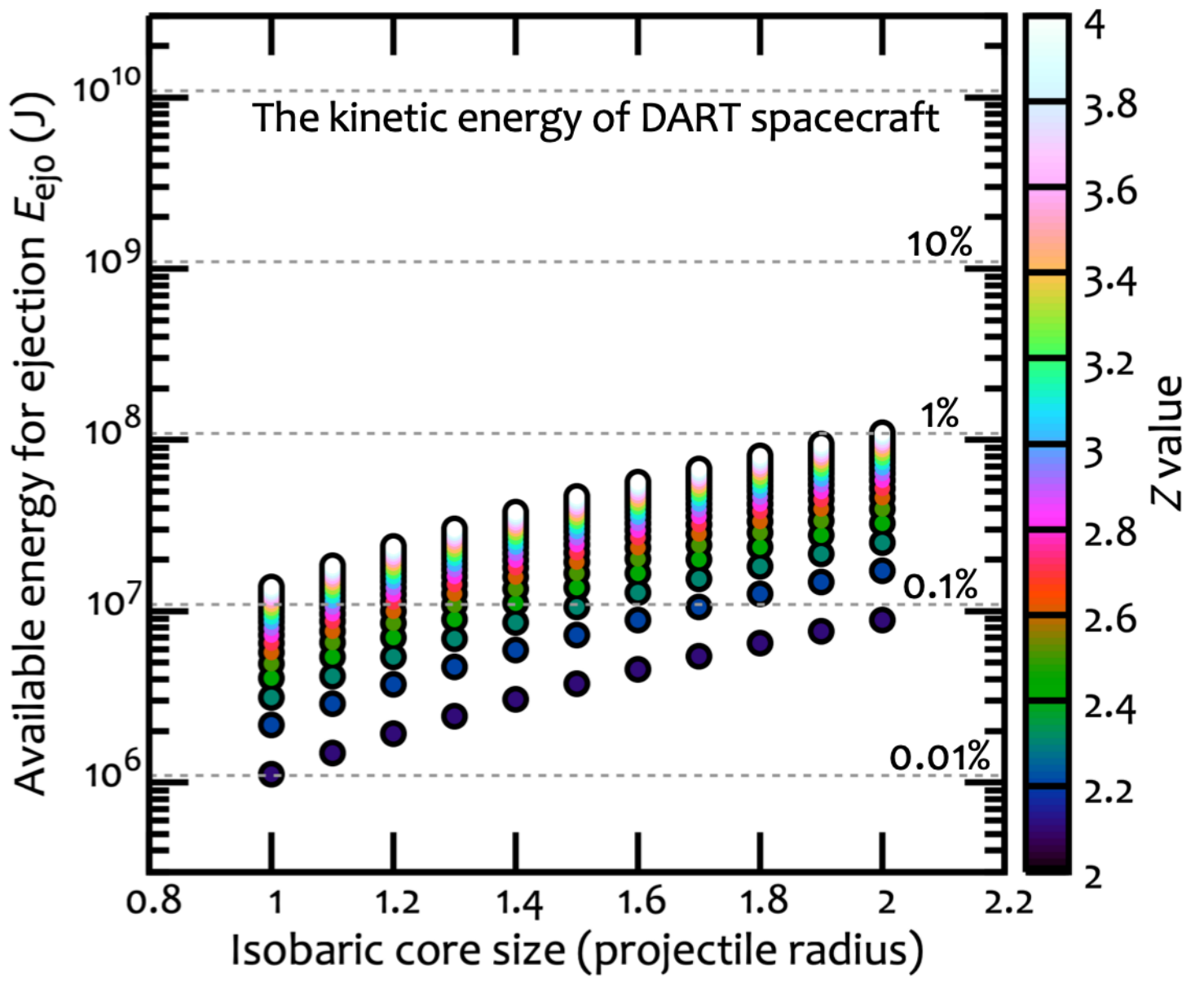

**Figure 6.** Energy budget pertaining to impact ejection as a function of the radius of the isobaric core calculated with (13). The color bar shows Maxwell $Z$ values. The kinetic energy of the DART spacecraft prior to impact and the fractions of this kinetic energy are shown as horizontal dashed lines.

## 3. RESULTS

In this section, we present the calculation results. We first applied the model to flat targets to compare our results with the predictions of the π-group scaling laws. In section 3.1, we show that our results are consistent with conventional crater scaling laws. In section 3.2, we show the results of a parameter survey. The total mass escaping from Dimorphos and the strength of the material on Dimorphos were constrained using the observed $\beta$ value from the DART collaboration team. In section 3.3, we show the results of in-depth analyses with the parameter set that yields reasonable $\beta$ values, including the transient crater profiles and ejecta mass–velocity distributions. In section 3.4, we discuss the globally averaged ejecta thickness on both Didymos and Dimorphos. Since we assumed a density of 2630 kg m$^{-3}$ as described in section 2.5, the expected $\beta$ range corresponds to 3.27–4.59 (3$\sigma$) for comparison with the observations (Cheng et al., 2023).

### 3.1. Model validation

The transient crater diameter $D_{\mathrm{tr}}$ under given impact conditions can be predicted using the π-group scaling laws (e.g., Holsapple and Schmidt, 1982; Housen and Schmidt, 1987), which were developed based on the late-stage equivalence concept (Dienes and Walsh, 1970) and Buckingham's $\pi$ theorem (Buckingham, 1914). The non-dimensional measures $\pi_{\mathrm{D}}$ and $\pi_2$ are defined as follows:

$$\pi_{\mathrm{D}} = \left(\frac{\rho}{M_{\mathrm{proj}}}\right)^{\frac{1}{3}} D_{\mathrm{tr}}, \quad (20)$$

and

$$\pi_2 = \left(\frac{4\pi}{3}\right)^{\frac{1}{3}} \left(\frac{gD_{\mathrm{proj}}}{v_{\mathrm{imp}}^2}\right), \quad (21)$$

where $M_{\mathrm{proj}}$ is the projectile mass and $D_{\mathrm{proj}}$ is the projectile diameter. $\pi_{\mathrm{D}}$ corresponds to the size ratio of the transient crater diameter to the projectile diameter, and $\pi_2$ corresponds to the ratio of gravitational to inertial stress (e.g., Melosh, 1989, P. 117). If target strength can be neglected (i.e., a gravity-dominated regime), then $\pi_{\mathrm{D}}$ can be simply obtained as a power law of $\pi_2$. For flat targets, twice the radial distance $R$ at which $E_{\mathrm{kin}} = E_{\mathrm{grav}} + E_{\mathrm{str}}$ corresponds to $D_{\mathrm{tr}}$. Figure 7 shows $\pi_{\mathrm{D}}$ as a function of $\pi_2$. We varied only the surface gravitational acceleration to change $\pi_2$, while keeping the projectile size and impact velocity fixed. The calculation results are shown along with two typical scaling lines for dry sand and saturated soil (Housen and Schmidt, 1987). The results for flat targets lie between the two scaling lines.

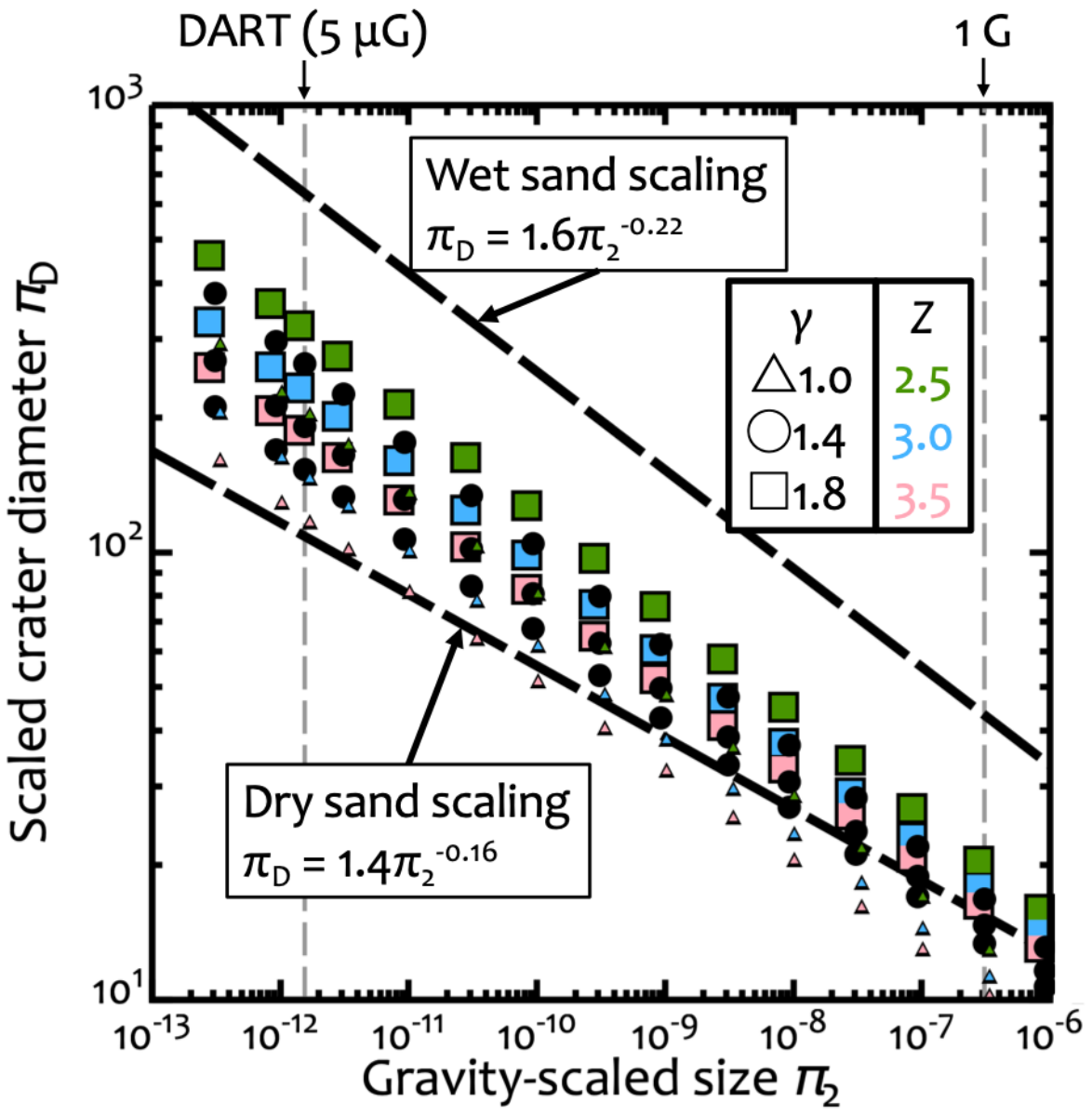

**Figure 7.** Relationship between the non-dimensional measures $\pi_D$ and $\pi_2$ in a gravity-dominated regime. The definitions of $\pi_D$ and $\pi_2$ are given by (20)–(21). We assumed flat and strengthless targets in this figure. The types of symbols and colors indicate the radius of the isobaric core normalized to the projectile radius, and the Maxwell $Z$ values used in the calculations, respectively. We used $Z = 2.5$, 3.0, and 3.5, as in Figs 1 and 3–4. For reasons of clarity, $\pi_2$ of each point is artificially shifted, depending on the $Z$ value. Two typical scaling lines with the equations are shown as dashed lines. The $\pi_2$ values for DART-like impacts under the gravitational conditions of Dimorphos (5 μG) and Earth (1G) are indicated by the grey dashed vertical lines.

Figure 8 shows the ejection velocity $v_{ej}$ as a function of the horizontal launch distance from the impact point $L_{ej}$. The results are well approximated with power laws down to a $v_{ej}$ of 0.1 m s$^{-1}$. According to the $\pi$-group scaling laws, the exponent is related to the velocity exponent $\mu$ (e.g., Housen et al., 1983), which ranges from 1/3 to 2/3 as $v_{ej} \propto L_{ej}^{-1/\mu}$. The slopes of our results on a log–log plot are within the allowable range of $\mu$. Consequently, our model reproduces the conventional results for flat targets with respect to both crater diameter and ejection behavior.

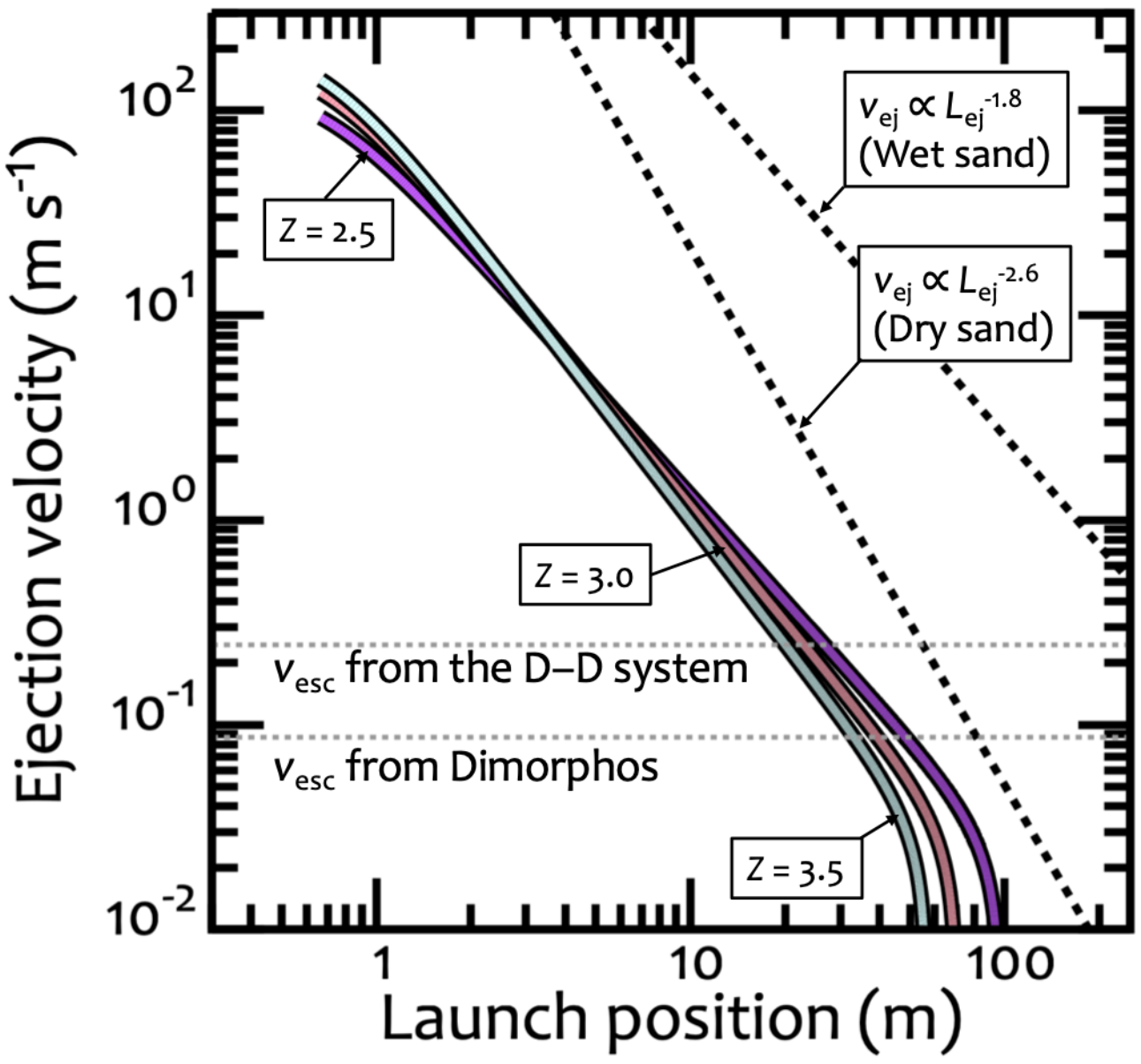

**Figure 8.** Ejecta velocity distributions as a function of the distance of the launch position from the impact point. We assumed flat and strengthless targets in this figure. We used $Z = 2.5$, 3.0, and 3.5, as in Figs 1, 3–4, and 7. The color indicates the ejection angle. Reference slopes corresponding to velocity exponents $\mu$ of 0.5 and 0.6 are shown as dotted lines. The escape velocities from the binary system (the D–D system) and from Dimorphos are also shown as horizontal dotted lines.

### 3.2. Material strength and the mass escaping from Dimorphos

By taking advantage of our quick-look tool, we conducted a parameter survey. We calculated the cumulative excavated mass escaping from Dimorphos and the $\beta$ values for both flat and spherical targets by changing the input parameters. Figure 9 shows $\beta$ as a function of constant strength $Y$. The dependence of $\beta$ on the Maxwell $Z$ value and isobaric core size ratio γ is also shown. We verified that $\beta$ is overestimated when flat targets are considered, as noted in previous studies (e.g., Raducan et al., 2023; Hirabayashi et al., 2025). Our model reproduced the target range of $\beta$ with $\gamma = 1.8$ and $Y < 10^2$ Pa in the case of a spherical target with a radius equal to that of Dimorphos. The momentum transfer is solely controlled by gravity at $Y \leq 0.1$ Pa. In general, the yield strength of rocky media depends on cohesion, which corresponds to the yield strength at zero pressure, the coefficient of friction, and pressure (e.g., Collins et al., 2004). Since the pressure field is expected to be intensely disturbed by several orders of magnitude during a shock-release cycle, the value of $Y$ also depends on time (e.g., Collins et al., 2004). With the assumed volumetric strain of unity, the effective strength parameter $\varepsilon Y$ in our model represents the energy dissipated per unit excavated volume. This energy dissipation includes contributions from plastic deformation and pore crushing in porous materials, although these processes are not explicitly modeled. Accordingly, $Y$ is treated as an input parameter (Table 1) rather than being calculated from a constitutive model.

Figure 10 shows $\beta$ as a function of the mass escaping from Dimorphos. Only the results for $\gamma = 1.8$ are shown in the figure. The $\beta$ value observed after the DART impact implies a significant mass loss from Dimorphos. A relatively low $Z$ value of $\leq 2.8$ is disfavored because a low $Z$ value results in relatively low ejection angles and inefficient momentum transfer. The escaping mass reaches 0.3%–3.0% of the mass of Dimorphos. The estimated strength and mass loss are consistent with the results obtained with full three-dimensional simulations (Raducan et al., 2024; Stickle et al., 2025).

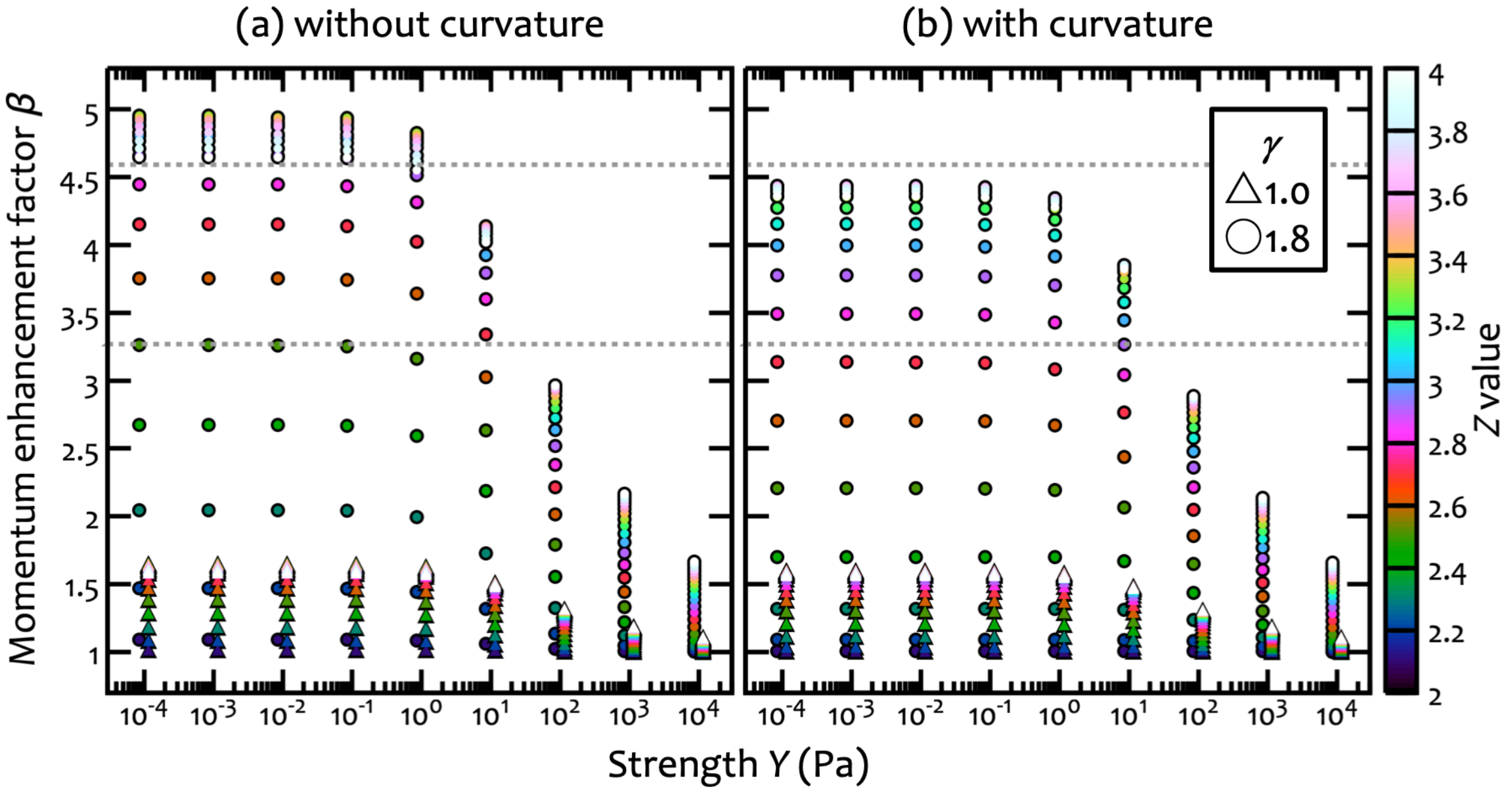

**Figure 9.** Momentum enhancement factor $\beta$ as a function of strength. The effects of target curvature, isobaric core size $\gamma$, and Maxwell $Z$ value on $\beta$ are also shown. For reasons of clarity, $Y$ of each point is artificially shifted, depending on the $\gamma$. The target values of $\beta$ (see the main text) are shown as two horizontal dashed lines.

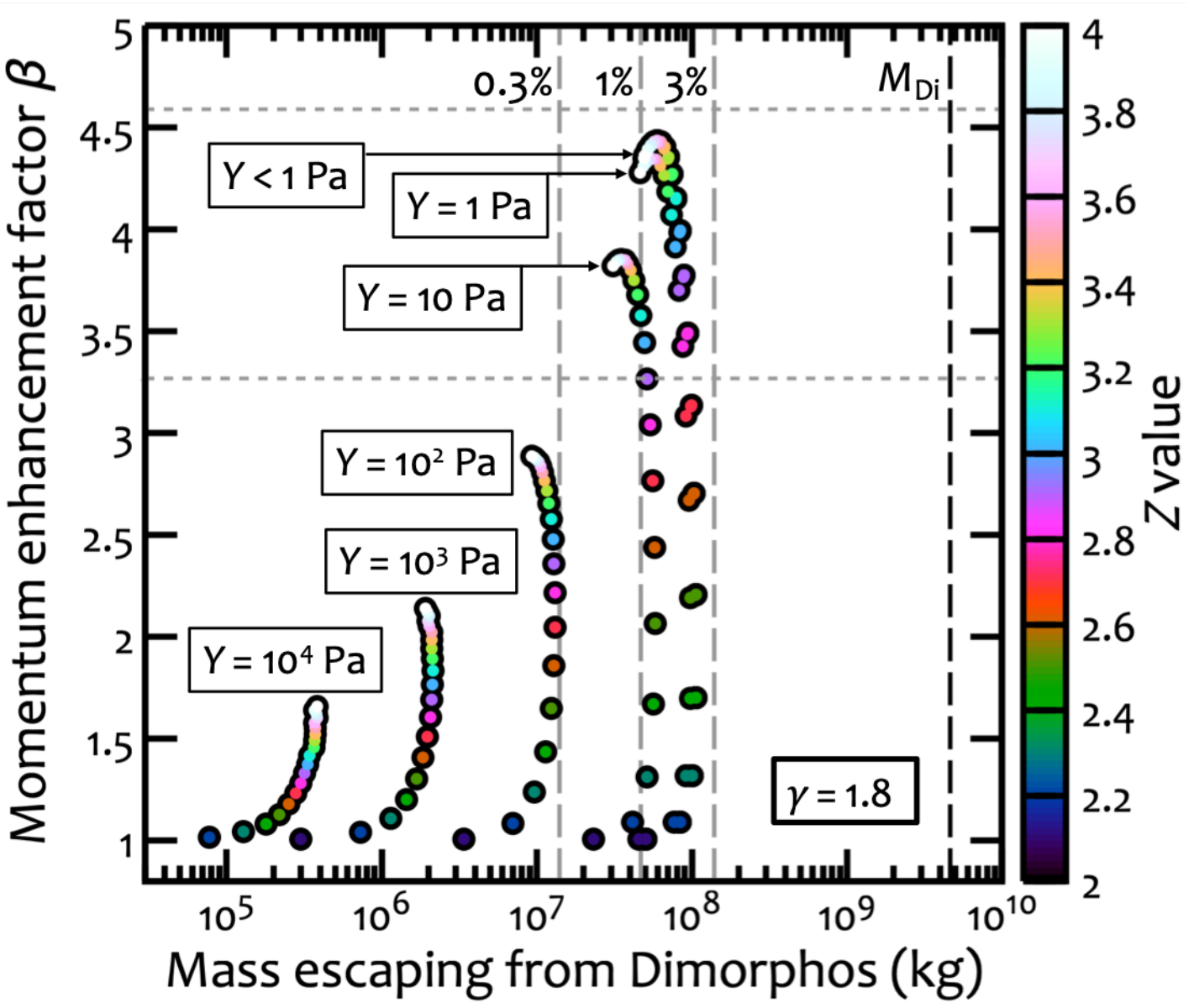

**Figure 10.** Relationship between the momentum enhancement factor $\beta$ and the mass escaping from Dimorphos. The color indicates the Maxwell $Z$ values used in the calculations. The isobaric core size was fixed at 1.8 × the projectile radius. The strength values $Y$ used in the calculations are indicated beside the data sequences pertaining to the same $Y$. The target values of $\beta$ are shown as two horizontal dashed lines as in Fig. 9. The four vertical dashed lines correspond to the mass of Dimorphos $M_{\mathrm{Di}}$ and fractions of this mass.

### 3.3. Transient crater profile and the ejecta mass–velocity distribution

In section 3.2, we constrained the parameters that yield $\beta$ values consistent with the DART observations. In this section, we mainly present the results for $Z = 3.0$, $Y = 10^{-2}$ Pa, and $\gamma = 1.8$, which yield $\beta = 4.0$ as a fiducial case. Here, $\gamma = 1.8$ is adopted as an effective energy-partition parameter, which reproduces the observed momentum-enhancement range, rather than as a direct estimate of the isobaric-core radius. For reference, we also plotted the results with $Z = 2.5$ and 3.5. The low strength $Y$ is equivalent to momentum transfer in a gravity-dominated regime as shown in Fig. 9. Figure 11 shows the cumulative excavated mass launched at a velocity faster than a given velocity $M_{\mathrm{exc}}$ ($>v_{\mathrm{ej}}$) as a function of $v_{\mathrm{ej}}$. The DART observations, including the continuous ejecta (Graykowski et al., 2023; Kim and Jewitt, 2023) and discrete boulders (Jewitt et al., 2023; Farnham et al., 2025), are also shown in the figure. Note that the ejecta masses were estimated by assuming specific material densities: 2400 kg m$^{-3}$ in Graykowski et al. (2023) and Farnham et al. (2025), 2200 kg m$^{-3}$ in Jewitt et al. (2023), and 3500 kg m$^{-3}$ in Kim and Jewitt (2023) although the ejection velocities were directly measured. To ensure a consistent comparison with our model, we recalculated the reported ejecta masses using the target density adopted in our calculations (2630 kg m$^{-3}$).

The fast component is beyond the scope of this study, because the speed is higher than the upper limit of the ejection velocity predicted by our model, which is the residual velocity in the isobaric core (Kurosawa and Takada, 2019). The upper limit of our model was calculated to be 360 m $s^{-1}$ with (11). In contrast, the speed of the fast ejecta was estimated to be 970 m $s^{-1}$ (Graykowski et al., 2023). This component may have been accelerated by different physical mechanisms, such as jetting (e.g., Melosh and Sonett, 1986; Johnson et al., 2014; Kurosawa et al., 2015) or spallation (Melosh, 1984; Polanskey and Ahrens, 1990; Kurosawa et al., 2018), rather than by the residual-velocity-driven excavation flow (Melosh, 1985; Kurosawa and Takada, 2019). Although the fast ejecta are not included in our model, the observed momentum enhancement factor $\beta$ can be reproduced by the ejecta launched by normal excavation alone under appropriate model conditions as shown in Figs. 9–10. Nevertheless, the fast ejecta may contribute to the total ejecta momentum, and their actual contribution remains uncertain.

The speed of the slow component is similar to our prediction. Our model allows us to track ejecta moving at a velocity much slower than the escape velocity. This capability is a key advantage of the proposed model. The mass–velocity relation deviates from a power-law behavior below $v_{esc}$ because gravitational potential energy becomes comparable to the kinetic energy of the ejecta. We can estimate (1) the mass of ejecta with initial velocities exceeding the escape velocity from the binary system $M_{esc,DD}$ ($v_{ej} > v_{esc,DD}$); (2) the mass injected into a dust torus around Didymos $M_{esc}$ ($v_{esc} < v_{ej} < v_{esc,DD}$); and (3) the mass deposited on Dimorphos $M_{dep}$ ($v_{ej} < v_{esc}$). We found that ~20% of the ejecta mass escaping from Dimorphos has initial ejection velocities exceeding the adopted escape velocity of the Didymos–Dimorphos system (24.5 cm $s^{-1}$), whereas the remaining ~80% has initial ejection velocities between the escape velocities of Dimorphos (8.7 cm $s^{-1}$) and the binary system. These fractions are based solely on the initial ejecta velocity distribution and do not necessarily represent the fractions that eventually escape from or remain gravitationally bound to the binary system.

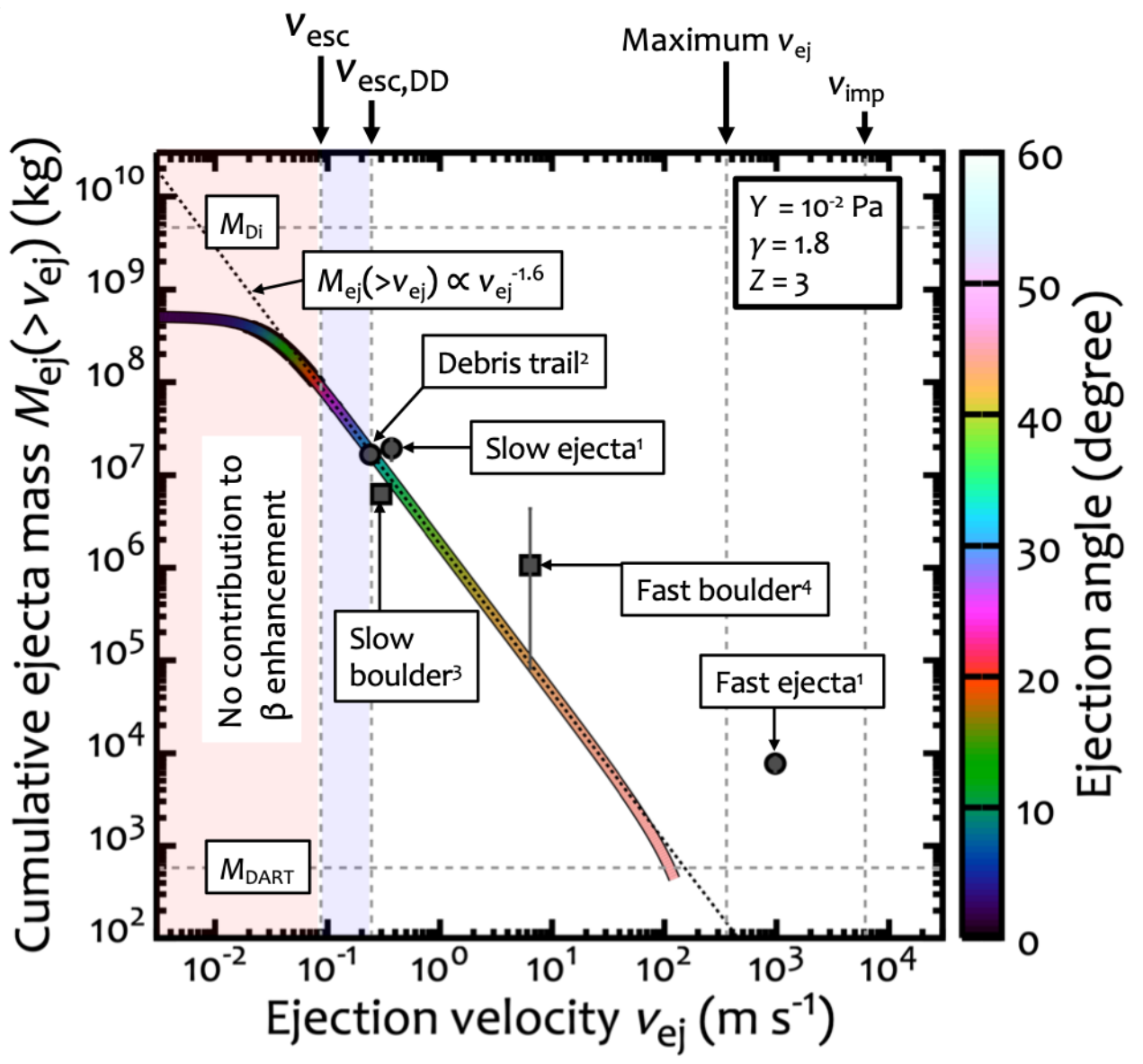

**Figure 11.** Ejecta mass–velocity distribution calculated with the selected parameters that are given in the figure. The color indicates ejection angle. The line with *Z* of 3.0 satisfies the observed momentum enhancement. The DART results pertaining to continuous ejecta ([1]Graykowski et al., 2023; [2]Kim and Jewitt, 2023) and discrete boulders ([3]Jewitt et al., 2023; [4]Farnham et al., 2025) are also shown as the filled circles and squares, respectively. The four vertical lines correspond to the impact velocity of the DART spacecraft, the upper limit of the ejection velocity in our model, the escape velocity from the binary system, and the escape velocity from Dimorphos (right to left). The blue and red shaded regions correspond to the masses of ejecta with initial velocities between the escape velocities of Dimorphos and the binary system and those with initial velocities below the escape velocity of Dimorphos, respectively.

Figure 12 shows the recoil momentum as a function of ejection velocity, demonstrating that ejecta over a broad range of ejection velocities contribute comparably to the total momentum transfer owing to the competing effects of ejection velocity, ejecta mass, and ejection angle.

Figure 13 shows a transient crater profile on a provenance plot. We also show the streamlines corresponding to the boundary between $M_{esc,DD}$ and $M_{esc}$, and between $M_{esc}$ and $M_{dep}$. The transient crater radius is estimated to be 60 m in this case, which is close to the radius of Dimorphos (75 m), suggesting that Dimorphos was substantially deformed as noted in previous studies (Raducan and Jutzi, 2022; Raducan et al., 2024; Daly et al., 2024; Naidu et al., 2024; Stickle et al., 2025). The maximum excavation depth measured from the surface can also be calculated as ~20 m. The ejecta excavated from the deepest part of the asteroid would be deposited near the crater rim.

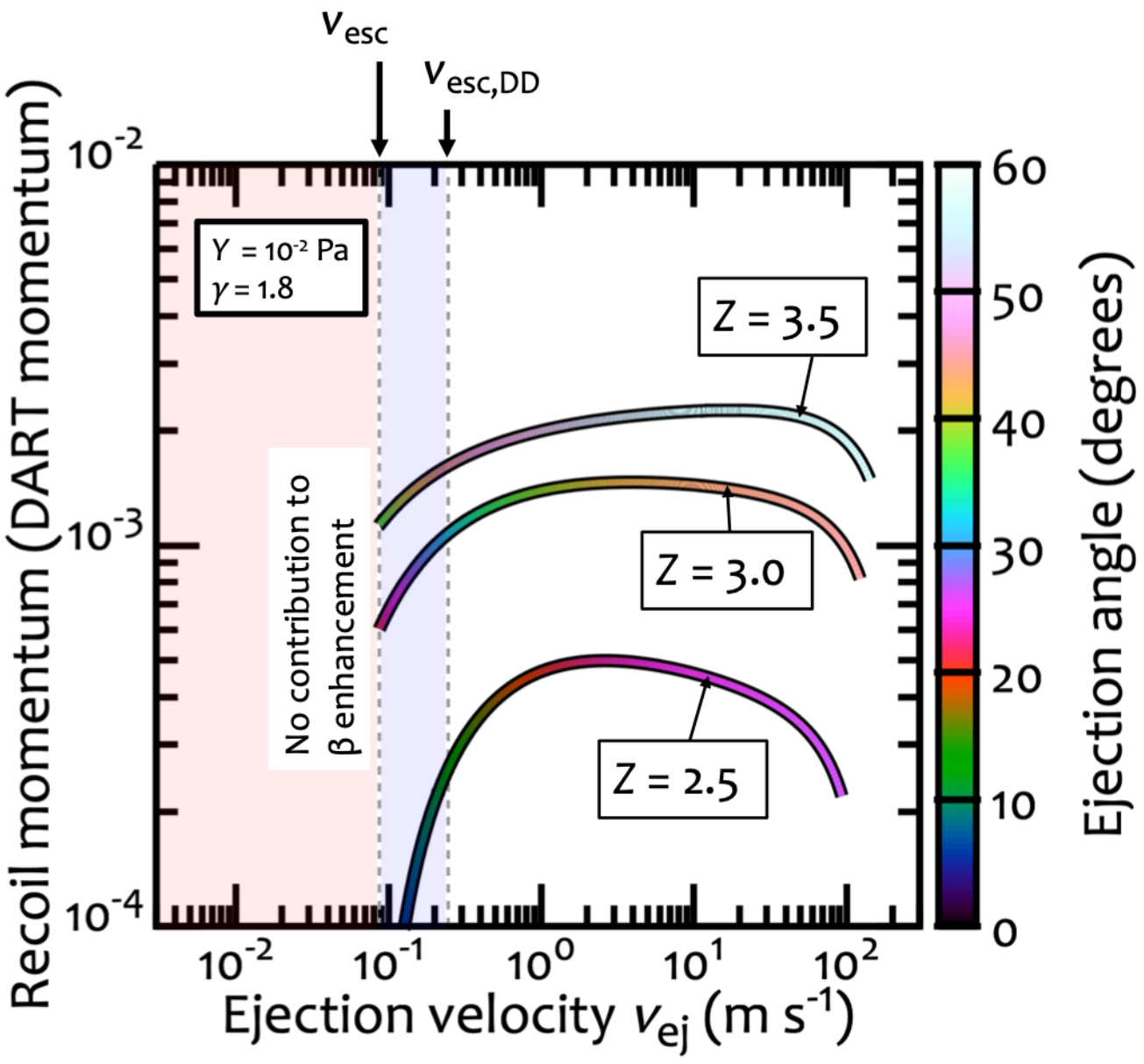

**Figure 12.** Recoil momentum as a function of ejection velocity. The same model parameters, shaded regions, and the color bar as those in Fig. 11 were used. Since the calculations were performed using a uniform radial increment ($\Delta R$ = 30% of the projectile radius), the corresponding velocity bins are not exactly uniform in width.

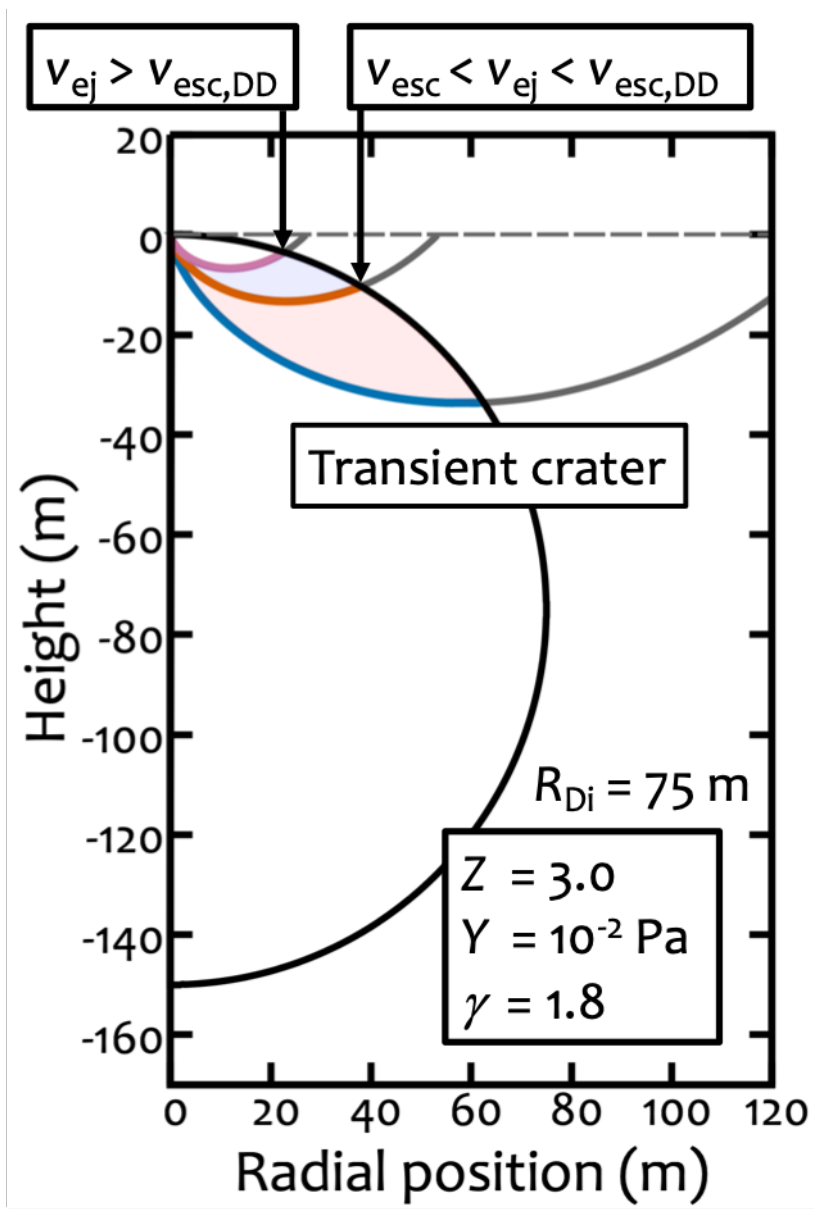

**Figure 13.** Transient crater profile, calculated with the parameters that reproduced the observed momentum enhancement, shown on a provenance plot. The same parameters as in Fig. 11 were used except that only the case with $Z$ = 3 is shown. The blue and red shaded regions correspond to the masses highlighted in Figs 11–12 with the same colors.

### 3.4. Ejecta thickness on Didymos and Dimorphos

We can estimate the globally averaged ejecta thickness $\overline{H_{ej}}$ on Dimorphos as follows:

$$\overline{H_{ej}} = \frac{M_{dep}}{\rho S_{Di}}, (22)$$

and

$$S_{Di} = 4\pi R_{Di}^2, (23)$$

where $S_{Di}$ is the surface area of Dimorphos. Given that we obtained $M_{dep} = 4 \times 10^8$ kg, which is ~10% of the mass of Dimorphos, $\overline{H_{ej}}$ reaches ~2 m.

Figure 11 shows that approximately $7 \times 10^7$ kg of Dimorphos material is ejected with initial velocities between the escape velocity of Dimorphos (8.7 cm $s^{-1}$) and the adopted escape velocity of the Didymos–Dimorphos system (24.5 cm $s^{-1}$). Some of these ejecta particles may enter orbits around Didymos and affect the post-impact orbital period of Dimorphos through angular momentum exchange (e.g., Naidu et al., 2024). Ejecta particles entering such orbits may subsequently be removed from the system by solar radiation pressure, as suggested by observations with the Hubble Space Telescope (Li et al., 2023).

As an illustrative estimate, if all $7 \times 10^7$ kg of this material were deposited uniformly over the surface of Didymos, the globally averaged ejecta deposit thickness would be approximately 15 mm, assuming a radius of 381 m for Didymos and a deposit density of 2630 kg $m^{-3}$. Although this scenario is highly idealized, the estimate provides a reference value for the potential scale of ejecta deposition on Didymos. The actual deposit thickness would depend on the subsequent dynamical evolution of the ejecta and their spatially nonuniform deposition, neither of which is considered in the present model.

Figure 14 shows the ejecta thickness distribution with the same parameters as in Figs 11–12. The ejecta thickness around a host crater decreases gradually with increasing distance from the crater rim as mentioned in previous studies (e.g., McGetchin et al., 1973). We found that the entire surface of Dimorphos is modified by the ejecta deposition, with an ejecta thickness of at least several tens of centimeters. The flight times of ejecta deposited at each site were calculated to be within 1 h, which is much shorter than the pre-impact rotation period of 11.7 h (Thomas et al., 2023). Recent SPH simulations by Raducan et al. (2026) also predict widespread ejecta re-accumulation on Dimorphos, with deposit thicknesses ranging from several tens of centimeters to more than 8 m. Despite differences in the modeling approaches and assumptions, both studies suggest that the DART impact caused substantial modification of the surface of Dimorphos through ejecta deposition.

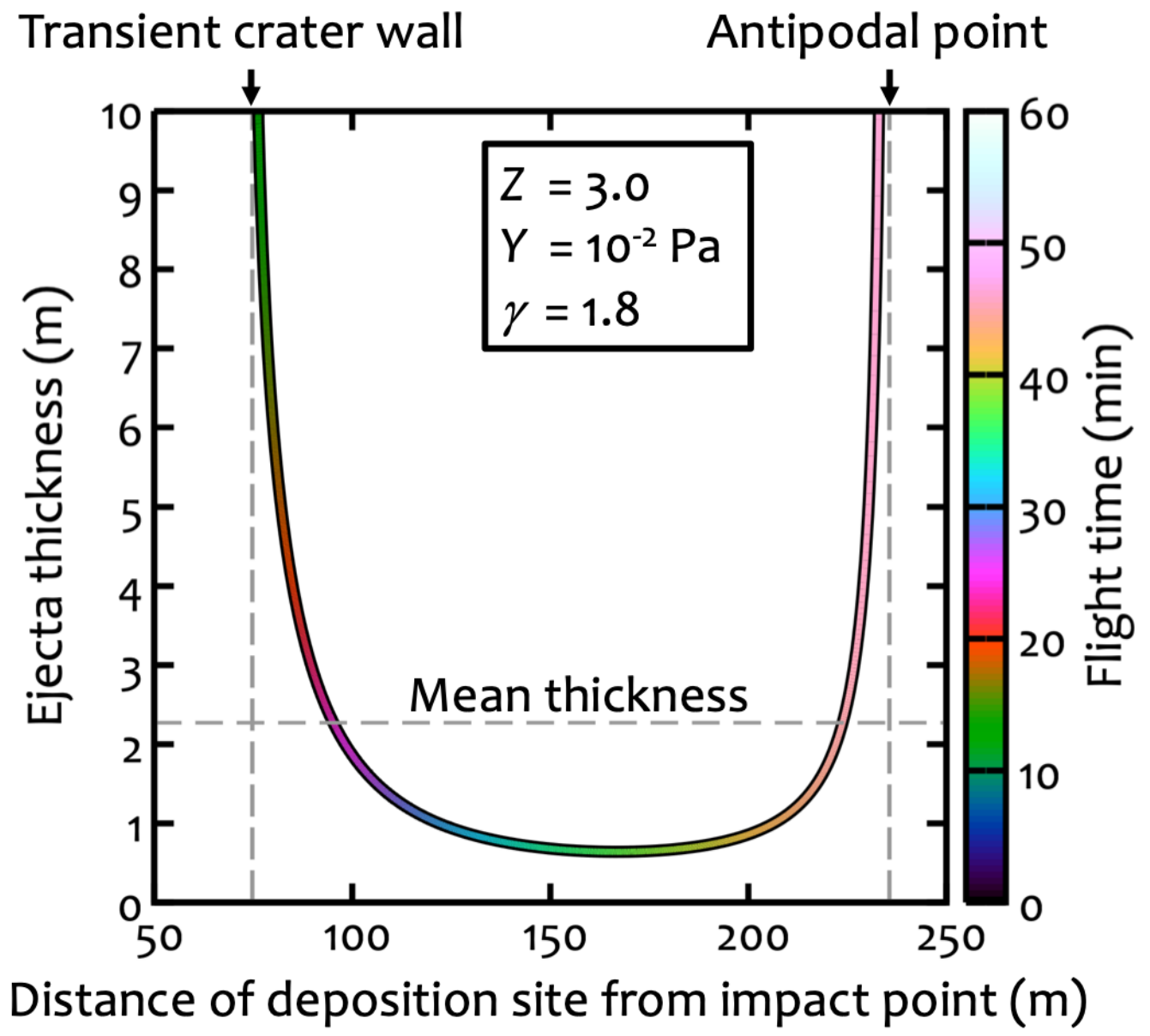

**Figure 14.** Distribution of ejecta thickness calculated with the parameters that reproduced the observed momentum enhancement. The parameters are the same as in Fig. 13. The horizontal dashed line is the globally averaged ejecta thickness calculated from (22)–(23). The two vertical lines correspond to the wall of the transient crater and the antipodal point of the impact. The color indicates the duration of ballistic flight.

## 4. DISCUSSION

In this section, we discuss the limitations of our model in section 4.1. Then, we propose Hera observational plans in section 4.2, and we discuss a lower bound of the ratio $\kappa_p$ and its implications for the density estimate of Dimorphos based on the sensitivity analysis conducted by Makadia et al. (2026).

### 4.1. Limitations of the present model

Maxwell's *Z* model has several inherent limitations, as discussed by Melosh (1989). In particular, the model neglects interactions between neighboring streamtubes, assumes a constant *Z* value although *Z* may vary with position and time (e.g., Austin et al., 1980), and does not accurately describe the excavation flow at distances much smaller or larger than the transient crater radius. These limitations also apply to our modified Maxwell's *Z* model (Kurosawa & Takada, 2019).

In addition, as mentioned in Section 2.2, we assume that the streamline geometry prescribed by Maxwell's *Z* model remains unchanged by target curvature. Although curvature is accounted for through the intersection of the streamlines with the target surface, its potential effects on the internal excavation-flow field are not explicitly considered. Hirabayashi et al. (2025) demonstrated that the Maxwell *Z* model can be used to interpret the three-dimensional geometry of the ejecta cone observed during the DART impact. However, the observed ejecta cone geometry is sensitive to the actual shape and local topography of the target surface. Because our model approximates Dimorphos as a perfect sphere and does not account for its irregular local topography, we did not use the observed ejecta curtain angle as an additional quantitative constraint on the model parameters.

Nevertheless, the model provides a useful analytical framework for estimating the overall characteristics of excavation flow without explicitly resolving the complex dynamics of crater formation. Therefore, our results should be regarded as approximate predictions of the ejecta mass and velocity distributions and their initial dynamical fate, rather than an exact description of the excavation flow.

### 4.2. The possibility of boulder burial by the ejecta blanket

The DRACO camera onboard the DART spacecraft captured boulders with sizes of 30 cm to 16 m (Barnouin et al., 2024; Pajola et al., 2024). The boulders on the pre-impact surface of Dimorphos are expected to be covered by the ejecta deposit, although some of the larger boulders may remain partially exposed. The degree of burial as a function of distance from the impact point, which will be investigated by Hera, might be useful for reconstructing the actual momentum transfer processes during the DART impact.

The ejecta thickness distribution shown in Fig. 14 is preliminary and represents only a first-order estimate, because we neglected deviations from a spherically symmetric gravitational field due to the irregular shape and possible internal density heterogeneity of Dimorphos, the rotation of Dimorphos, the orbital motion of Dimorphos around Didymos, and the gravitational influence of Didymos. Although these effects should be considered to accurately estimate the ejecta thickness distribution on Dimorphos, such advanced calculations are beyond the scope of the present study.

### 4.3. Re-evaluation of the mass and the density of Dimorphos

As shown in Figure 11, our model predicts that ~80% of the ejecta mass escaping from Dimorphos has initial ejection velocities below the adopted escape velocity of the Didymos–Dimorphos system under plausible impact conditions. This estimate does not account for the orbital velocity of Dimorphos, subsequent gravitational interactions within the binary system, or solar radiation pressure. Therefore, the actual fraction of ejecta that remains gravitationally bound to the system cannot be determined from this estimate alone. Nevertheless, our model provides an estimate of the ejecta momentum ratio $\kappa_p$ from the initial ejecta mass–velocity–angle distribution using the two adopted escape-speed thresholds.

Figure 15 shows $\kappa_p$ as a function of $\beta$. Our model predicts that $\kappa_p$ can be as low as approximately 0.6 within the plausible range of $\beta$, based on the initial ejecta distribution. However, the effective value of $\kappa_p$ associated with the observed dynamical response of the binary system may differ from this initial estimate because of the subsequent dynamical evolution of ejecta. For example, some ejecta initially classified as remaining within the binary system may eventually escape, potentially modifying the net momentum transferred to the system.

To explore the implications of our model prediction for the bulk density of Dimorphos, we considered an alternative uniform distribution of $\kappa_p \sim U[0.6, 1.0]$. This distribution is motivated by our model prediction but is not directly derived from it. By linearly interpolating the sensitivity analysis reported by Makadia et al. (2026) for uniform distributions of $\kappa_p \sim U[0.5, 1.0]$ and $\kappa_p \sim U[0.75, 1.0]$ (their Table S4), we estimated the bulk density of Dimorphos to be approximately 1700 ± 300 kg m$^{-3}$. Using the volume derived from the shape model of Dimorphos (Daly et al., 2024), the corresponding mass is approximately (3.0 ± 0.5) × $10^9$ kg. These estimates are approximate because they are based on interpolation of the published sensitivity analysis rather than a direct reanalysis of the observational data.

The estimated density is close to that of the other S-type asteroids, including Itokawa and Moshup. Nevertheless, it remains consistent with the nominal density estimate of 1540 ± 220 kg m$^{-3}$ reported by Makadia et al. (2026) when the associated uncertainties are taken into account. Although our calculations assumed a target density of 2630 kg m$^{-3}$ as mentioned in section 2.5, the inferred value of $\kappa_p$ is independent of the assumed density when the ejecta velocity distribution is held fixed, because the momentum terms in both the numerator and denominator of the ratio scale linearly with density. Therefore, within the present model, no iterative recalculation is required solely to rescale the ejecta momentum terms for a different target density. Our results can help inform the range of $\kappa_p$ values considered when interpreting the dynamical response to the DART impact. Detailed simulations of the subsequent ejecta dynamics, together with Hera observations, will be essential for further constraining $\kappa_p$ and refining the inferred mass and bulk density of Dimorphos.

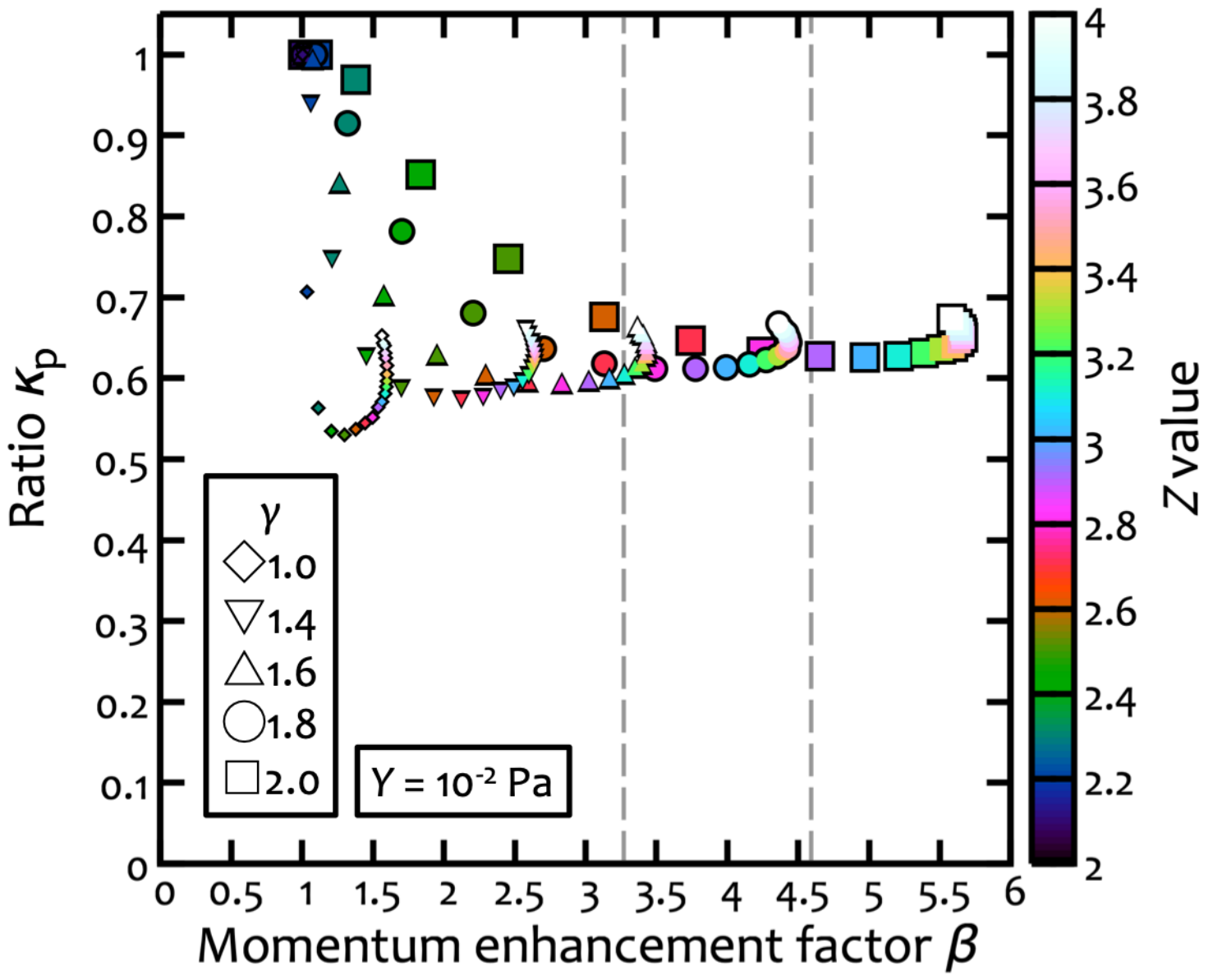

**Figure 15.** The ratio $\kappa_p$ as a function of $\beta$. Here, we used $Y = 10^{-2}$ Pa as in Figs. 11–14. The effects of isobaric core size ratio $\gamma$ and Maxwell $Z$ value on $\kappa_p$ are also shown. the observational bounds on $\beta$ are shown as two vertical dashed lines.

## 5. CONCLUSIONS

We developed a simple model to predict the outcomes of the DART impact based on a modified version of Maxwell's $Z$ model. Unlike conventional numerical models constrained by computational cost, our model enables efficient prediction of ejecta with velocities ranging from ~100 m s$^{-1}$ to below the escape velocity of Dimorphos. The model outputs are consistent with previous results, including predictions based on conventional crater scaling laws for flat targets, the strength of the material on Dimorphos, and the mass escaping from Dimorphos due to the DART impact. The agreement supports the validity of previous full three-dimensional shock physics simulations. The model also enables the investigation of the fate of relatively low-speed ejecta. The predicted mass and velocity distributions of the slow ejecta are broadly consistent with post-impact observations. We found that a substantial fraction (~80%) of the ejecta mass escaping from Dimorphos has initial ejection velocities between the escape velocity of Dimorphos and the adopted escape velocity of the Didymos–Dimorphos system. This fraction does not necessarily represent the ejecta mass that remains gravitationally bound to the binary system over long timescales. The maximum excavated depth beneath the surface during the DART impact was calculated to be ~20 m. The globally averaged and minimum ejecta thicknesses were estimated to be ~2 and ~0.5 m, respectively. A globally-averaged ejecta thickness on Didymos was estimated to be ~15 mm based on an optimistic assumption. We propose that the degree of burial of boulders as a function of distance from the impact point on Dimorphos could be used to reconstruct the momentum transfer processes driven by the DART impact. By combining our model with the recently detected heliocentric deflection of the system, we suggest that the bulk density of Dimorphos could be higher than the nominal estimate.

**Declaration of Competing Interests**

The authors declare no competing interests.

**Data availability**

The numerical data used to generate the figures presented in this study, together with the corresponding gnuplot scripts, are publicly available on Zenodo at https://doi.org/10.5281/zenodo.23943564. The dataset includes instructions for reproducing the figures.

**Acknowledgements**

We appreciate anonymous reviewers for constructive comments, and Kathryn Volk for handling the manuscript as the Scientific Editor. We thank Shota Kikuchi and Yuki Koyano for their helpful comments. K.K. is supported by JSPS KAKENHI Grants JP25K01063 and JP25H00692. S.T. is supported by JSPS KAKENHI Grants JP20K14428 and JP21H01006.

## Appendix A. Derivation of the volume element of an excavation streamtube Eq. (5)

In the Maxwell *Z* model, the excavation streamlines are described in spherical coordinates by (3) in the main text. Each streamline is identified by a constant value of *R*. Aassuming axial symmetry, the differential volume element d*V* in spherical coordinates ($r$, $\theta$) is given by

$$dV = 2\pi r^2 \sin\theta \, dr d\theta. \qquad \text{(A1)}$$

To express this volume element in terms of the streamline coordinates ($R$, $\theta$), we introduce the coordinate transformation $r = r(R, \theta)$. The corresponding Jacobian determinant is

$$J(R, \theta) = \left(\frac{\partial r}{\partial R}\right)_\theta . \qquad \text{(A2)}$$

From (3) and (A2),

$$J(R, \theta) = (1 - \cos\theta)^{\frac{1}{Z-2}}. \qquad \text{(A3)}$$

Therefore, the volume element in the streamline coordinates becomes

$$dV = 2\pi R^2 (1 - \cos\theta)^{\frac{3}{Z-2}} \sin\theta \, dR d\theta. \qquad \text{(A4)}$$

The derived (A4) is equivalent to (5) in the main text and is used to calculate the excavated mass and the energy contributions associated with each streamtube in our model.

## Appendix B. Derivation of the depth *H* below the surface at a point along a streamline

In our model, the depth *H* below the surface along each streamline is required to evaluate the integral in Equation (7). Figure A1 illustrates the geometry considered in our model, showing the surface of Dimorphos and an example of an excavation streamline. Applying the law of cosines to the triangle defined by $R_{Di}$, $r$, and ($R_{Di} - H$), we obtain

$$H = R_{Di} \left[ \left( 1 - \sqrt{1 + \left(\frac{r}{R_{Di}}\right)^2 - 2\left(\frac{r}{R_{Di}}\right)\cos\theta} \right) \right]. \qquad \text{(B1)}$$

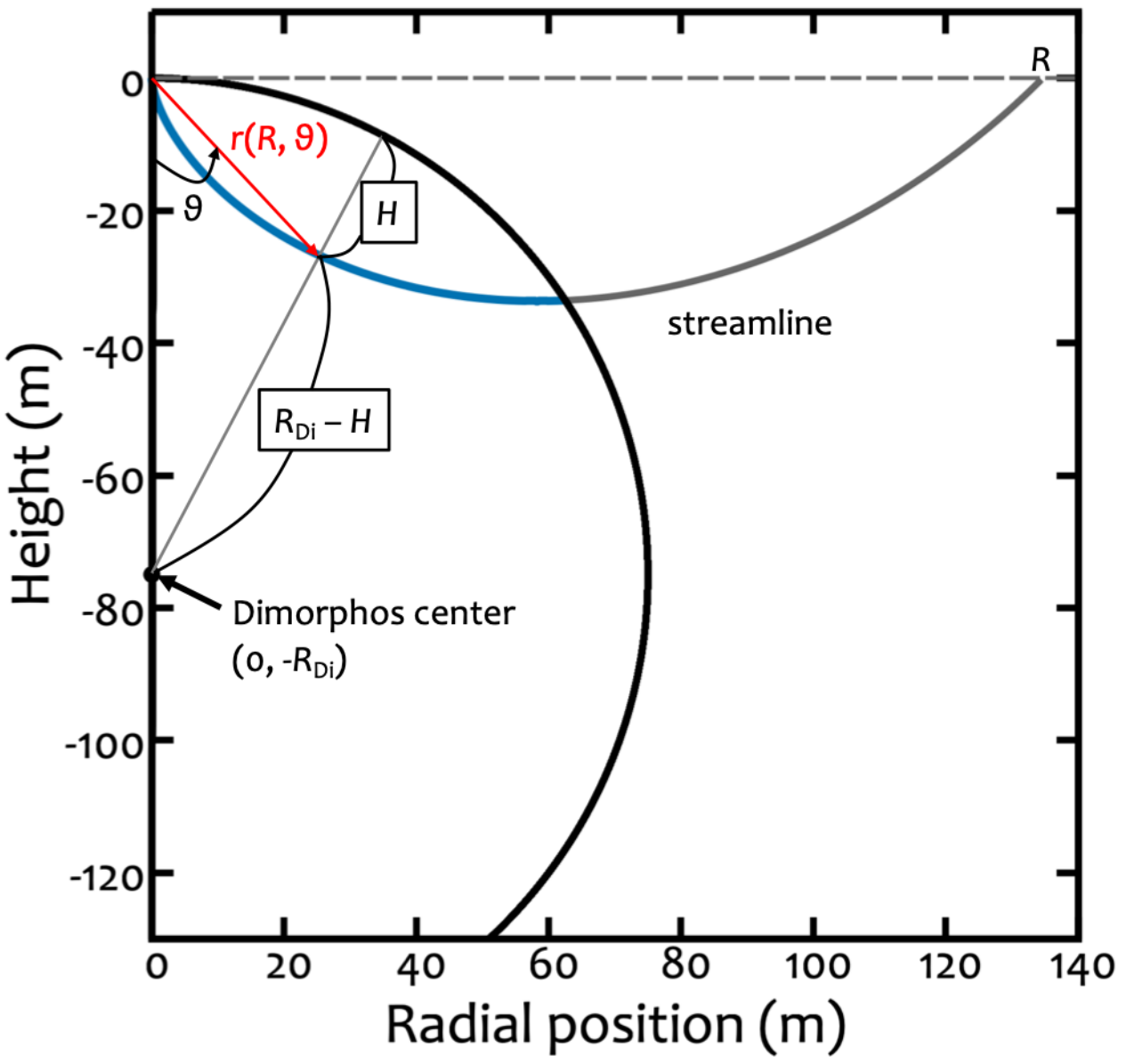

**Figure B1.** Schematic illustration of the geometry used to define the depth $H$ below the surface of Dimorphos in two-dimensional polar coordinates.

**Appendix C. The derivation of $M_{hem}$ defined in Figure 5 and used in (14)**

In Section 2.3, we introduced $M_{hem}(R)$, the mass of the region defined in Equation (14), as a function of the radial distance $R$. Figure C1 illustrates the geometry used to calculate $M_{hem}(R)$. We consider both Cartesian and polar coordinate systems in two dimensions. The surface of Dimorphos is expressed in Cartesian coordinates as

$$x^2 + (y - R_{\mathrm{Di}})^2 = R_{\mathrm{Di}}^2. \qquad \text{(C1)}$$

The vertical distance $\Delta y(x)$ between the surface of Dimorphos and the hemisphere of radius $r_c$, at a given horizontal coordinate $x$, is given by

$$\Delta y(x) = r_c \cos\theta - R_{\mathrm{Di}} + \sqrt{R_{\mathrm{Di}}^2 - x^2}. \qquad \text{(C2)}$$

Here, $r_c$ is the radius of the hemisphere illustrated in Figure 5 of the main text. This radius corresponds to the distance from the origin to the intersection of a streamline with the surface of Dimorphos at a given $R$. Thus, $r_c = r(R, \theta_c)$, where $r(R, \theta_c)$ is calculated using (3), and $\theta_c$ is defined by (4). Assuming axial symmetry, the mass $M_{hem}(R)$ can be calculated by integrating the volume elements over the region of interest as

$$M_{\mathrm{hem}}(R) = 2\pi\rho \int_0^{r_c \sin\theta_c} x \Delta y \mathrm{d}x. \tag{C3}$$

Using the substitutions $x = r_c \sin\theta$ and $\mathrm{d}x = r_c \cos\theta \mathrm{d}\theta$, we obtain

$$M_{\mathrm{hem}}(R) = 2\pi\rho \int_0^{\theta_c} r_c^2 \sin\theta\cos\theta \left( r_c\cos\theta - R_{\mathrm{Di}} + \sqrt{R_{\mathrm{Di}}^2 - r_c^2 \sin^2\theta} \right) \mathrm{d}\theta. \tag{C4}$$

In our calculations, (C4) was evaluated numerically at each integration step with respect to $R$.

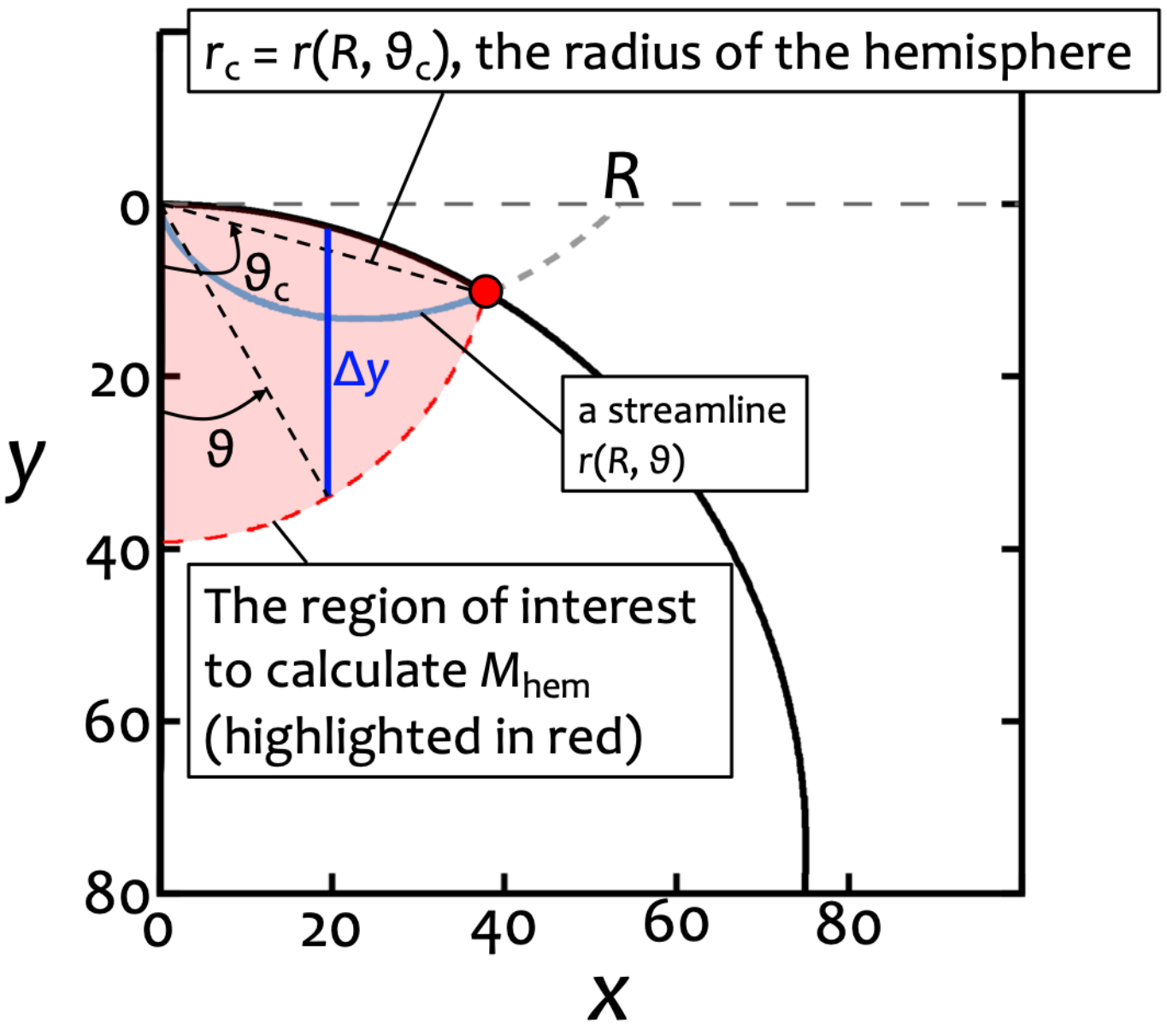

**Figure C1.** Geometry and dimensions used to calculate the mass $M_{\mathrm{hem}}(R)$ of the region of interest (highlighted in red).

## Appendix D. Resolution dependence

As described in Section 2.4, our model involves numerical integration over two variables, $R$ and $\theta$. To assess the effects of numerical resolution on the calculated momentum enhancement factor $\beta$, we performed a resolution sensitivity analysis by varying the integration step sizes, $\Delta R$ and $\Delta\theta$, under a representative set of impact conditions that reproduces the observed $\beta$.

Figure D1 shows that the calculated $\beta$ exhibits an approximately linear dependence on $\Delta R$ for $\Delta R < R_{\mathrm{proj}}$. By extrapolating this linear relationship to ($\Delta R = 0$), we estimate the value of $\beta$ in the limit of infinite resolution. We confirmed that the calculated $\beta$ obtained with ($\Delta R = 0.3R_{\mathrm{proj}}$), which was adopted in the parameter survey conducted in this study, differs from the extrapolated value by less than 5%.

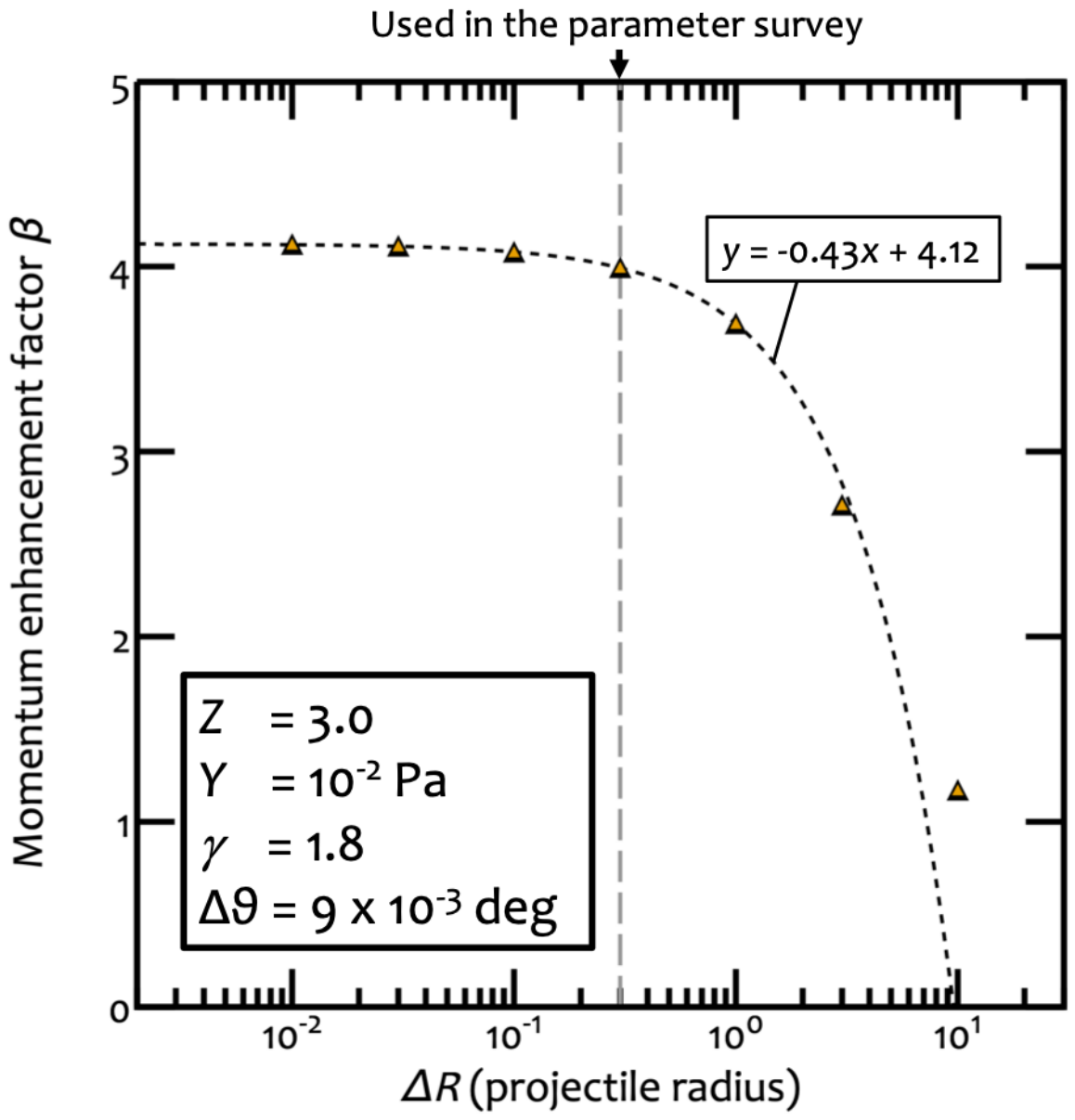

**Figure D1.** Dependence of the calculated momentum enhancement factor $\beta$ on the radial integration step size $\Delta R$. The calculation conditions, linear regression line, and corresponding fitting equation are shown in the figure.

Figure D2 shows that the dependence of the calculated $\beta$ on $\Delta\theta$ is substantially weaker than its dependence on $\Delta R$. The adopted angular integration step size of $\Delta\theta = 9 \times 10^{-3}$ degrees provides a sufficiently converged value of $\beta$.

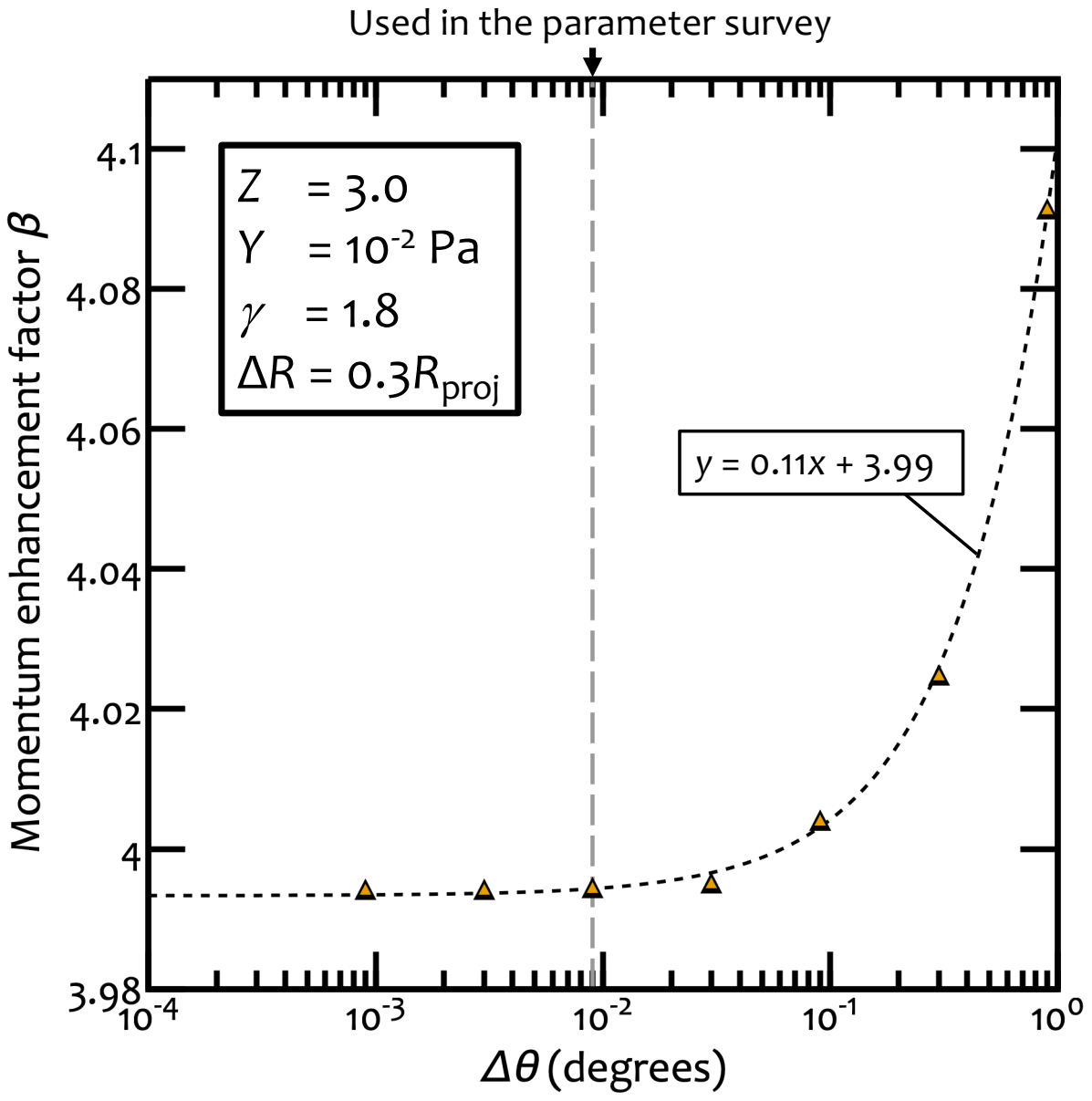

**Figure D2.** Same as Figure D1, but for the angular integration step size $\Delta\theta$.